\documentclass[a4paper,11pt]{article}
\usepackage[utf8]{inputenc}
\usepackage{color}
\usepackage{cite}
\usepackage{hyperref}
\usepackage{amsmath}
\usepackage{graphicx}

\usepackage[top=0.4in, bottom=0.7in, left=0.60in, right=0.5in]{geometry}

\begin{document}
\title{{\Large ($\gamma$, $\chi_{cJ}(J=0,1)$) and ($\gamma$, $\eta_c$) production in electron-positron annihilation at $\sqrt{s}=10.6$ GeV and $4.6$ GeV in the framework of Bethe-Salpeter equation}}
\author{Shashank Bhatnagar, Vaishali Guleria}
\maketitle \small{Department of Physics, University Institute of Sciences, Chandigarh University, Mohali-140413, India\\}

\begin{abstract}
\normalsize{In present work we study the production of ground and excited charmonium states in $e^- e^+ \rightarrow \gamma\chi_{cJ}(nP)(J=0,1)$, and  $e^- e^+ \rightarrow \gamma \eta_c(nS)$,  through leading order (LO) diagrams, which proceed through exchange of a virtual photon that couples to $\gamma$ and $\eta_c/\chi_{cJ}$ through the triangular quark loop diagram, in the framework of $4\times 4$ Bethe-Salpeter equation (BSE), at center of mass energies  $\sqrt{s}=10.6$ GeV (Belle energy), and $4.6$ GeV (BESIII energy). The amplitude simplifies to a general form required by Lorentz-covariance, in terms of its form factors. The cross sections for these processes with leading order diagrams alone at $\sqrt{s}=10.6$ GeV provide a sizable contribution, which might be mainly due to the BSE being a fully relativistic approach that incorporates the relativistic effect of quark spins and can also describe internal motion of constituent quarks within the hadron in a relativistically covariant manner. Our results are compared with recent Belle data at 10.6GeV, and BESIII data at 4.6GeV, as well as other models. Plots of cross sections versus the center-of mass energy, $\sqrt{s}$ reveal mild fluctuations in their behaviour for all the three processes in the low energy ($\sqrt{s}$ = 4 - 6 GeV) region, and are analyzed in terms of the form factors.}
\end{abstract}

\section{Introduction}
Quarkonium production processes are of theoretical and experimental interest. However, there was a significant discrepancy between the experimental measurements by BABAR and Belle collaborations\cite{belle10,babar09,ulgov05} of total cross sections of $e^- e^+ \rightarrow \gamma \rightarrow J/\Psi \eta_c$ and $e^- e^+ \rightarrow \gamma \rightarrow J/\Psi \chi_{cJ}$ at energy, $\sqrt{s}=10.6$ GeV, and their NRQCD predictions\cite{braaten,bodwin} which were an order of magnitude smaller than experiment. Interestingly, the leading order QCD diagrams alone gave cross-sections which were much less than data. And incorporation of higher order QCD corrections\cite{zhang05} could increase the cross section only by a factor of around $1.8$, but this was again insufficient to reach the experimental results of BABAR and Belle.

A number of studies were carried out to resolve this problem. In an interesting study, it was recently pointed out that for both the processes, $e^- e^+ \rightarrow J/\Psi \eta_c$ and $e^- e^+ \rightarrow \gamma \rightarrow J/\Psi \chi_{cJ}$  in addition to the NLO QCD corrections, the interference between the QCD and QED tree-level diagrams\cite{sun18,sun18a}, can also provide signiﬁcant contributions. Their studies have further shown that the colour octet contributions to $e^- e^+\rightarrow \gamma \rightarrow J/\Psi \eta_c$, and $e^- e^+ \rightarrow \gamma \rightarrow J/\Psi \chi_{cJ}$ are negligibly small, with the dominant contributions \cite{sun18a,sun18} coming from the colour singlet channels.

Further, it was also found that the discrepancy between experimental results and theoretical predictions for the cross section of $e^- e^+\rightarrow J/\Psi \eta_c$ could also be resolved by taking into account the intrinsic motion of quarks inside the produced hadrons in the framework of light-cone expansion method\cite{braguta05,luchinsky}, which NRQCD does not take into account at leading order.

Now, there are also processes, $e^-+e^+\rightarrow \gamma +H$, that proceed through a virtual photon, where $H$ is a heavy quarkonium of charge conjugation parity, $C=+1$. Here, $H$ can be S-wave spin singlet states such as $\eta_c(nS),\eta_b(nS)$, or P-wave spin triplet states such as, $\chi_{cJ}$ and $\chi_{bJ}$, with $(J=0,1,2)$. And very recently, the experimental study of the $e^-e^+\rightarrow \gamma \chi_{c0,1,2}$\cite{ablikim21} processes was carried at center-of-mass energies ranging from $\sqrt{s}=4.008-4.6$ GeV. using data samples corresponding to an integrated luminosity of $16 fb^{-1}$ accumulated with the BESIII detector at the BEPCII collider. The production cross sections of $e^-e^+\rightarrow \gamma\chi_{c0,c1}$ at each center of mass energy was measured, and the cross section for these processes, $\sigma(e^-e^+\rightarrow \gamma\chi_{c0})=2.6\times 10^{3}$ fb, and $\sigma(e^-e^+\rightarrow \gamma\chi_{c1})=(1.7^{+0.8}_{-0.6}\pm 0.2)\times 10^{3}$ fb at $\sqrt{s}=4.5995$ GeV, using integrated luminosity of $19.3 fb^{-1}$ was reported by BESIII at BEPCII. The other experimental studies on $e^-e^+\rightarrow \gamma\chi_{c1,c2}/\eta_c$ above 4 GeV, carried out before this study were by BESIII\cite{bes15,ablikim17}, CLEO\cite{cleo06} and Belle\cite{belle15}. A more recent experimental study on cross sections for $e^-e^+\rightarrow \gamma\chi_{c0,c1,c2}$ and $e^-e^+\rightarrow \gamma\eta_c$ was carried out by Belle\cite{jia18} at energies, $\sqrt{s}=10.52$ GeV, $10.58$ GeV, and $10.867$ GeV.  At $\sqrt{s}=10.58$ GeV, these cross sections\cite{jia18} were, $\sigma(e^-e^+\rightarrow \gamma\chi_{c0})=-20.0^{+122.3}_{-111.0}\pm2.6$ fb, $\sigma(e^-e^+\rightarrow \gamma\chi_{c1})=17.3^{+4.2}_{-3.9}\pm1.7$ fb, and $\sigma(e^-e^+\rightarrow \gamma\eta_{c})=11.3^{+7.0}_{-6.6}\pm 1.5$ fb.

Some of the recent works have addressed exclusive production of $\chi_{cJ}\gamma$ in $e^- e^+$ collisions in the NRQCD factorization framework\cite{brambilla17}, with calculations of leading relativistic corrections, and the estimates for the cross sections made. Further, pNRQCD predictions\cite{brambilla20} for exclusive electromagnetic production of $\chi_{cJ}\gamma$, at 10.6 GeV, and $\chi_{bJ}\gamma$ in the energy range, 20 - 500 GeV, that encompasses the energies of possible future $e^- e^+$ collider. In another recent work\cite{sang20}, within the NRQCD factorization framework, authors calculated the $O(\alpha_{s}^2)$ corrections to exclusive production of $\chi_{cJ}\gamma$ at B-factories, and have verified the validity of NRQCD factorization for exclusive P-wave quarkonium production to two-loop order, where $O(\alpha_{s}^2)$ corrections were found to be smaller than the $O(\alpha_s)$ corrections for all the three channels.

In this work we treat the processes, $e^-+e^+\rightarrow \gamma^*\rightarrow \gamma +H$ using the framework of Bethe-Salpeter equation (BSE) \cite{shi21,yu19, chang10, guo08, mitra01, eshete19,bhatnagar20,vaishali21,wang22}, since BSE is a very useful non-perturbative tool for studying relativistic bound state problems. Due to its firm base in quantum field theory and being a dynamical equation based approach, it provides a realistic description for analyzing hadrons as composite objects, and is a fully relativistic approach that incorporates the relativistic effect of quark spins, and can also describe internal motion of constituent quarks within the hadron in a relativistically covariant manner. Thus, it can be applied to study processes over a wide range of energies, such as the mass spectrum of mesons, their form factors, decays, and also their productions, due to its Lorentzian covariance  and relativistic character. Thus its application to charmonium production processes is of interest. Thus, in this work, we start by calculating the cross section for the three processes,  $e^- + e^+ \rightarrow \gamma + \chi_{cJ} (J=0,1)$, and $e^- + e^+ \rightarrow \gamma + \eta_c$, taking only the leading order (tree-level) diagrams $\sim O(\alpha_{em}^3\alpha_s^0)$ in QED and QCD at two center of mass energies, $\sqrt{s}$=10.6 GeV.(corresponding to Belle) and 4.6 GeV.(corresponding to BESIII), and our results are then compared with recent data from both Belle\cite{jia18}, and BESIII\cite{ablikim21}. These processes proceed through a virtual photon, that is in turn coupled to $\gamma$ and $\chi_{cJ}/\eta_c$ through the quark-triangle diagrams shown in Fig.1. In this work we focus only on the colour singlet contributions \cite{sun18}, and derive analytical expressions for their cross sections in the framework of Bethe-Salpeter equation.

To calculate the above mentioned processes, we will make use of the calculational techniques involved in the quark-triangle diagrams that were recently used for the study of $M1$ and $E1$ radiative transitions involving heavy-light quarkonia, such as: $1^{--}\rightarrow 0^{-+}\gamma$, $0^{-+}\rightarrow 1^{--}\gamma$\cite{bhatnagar20,vaishali21}, and $1^{+-}\rightarrow 0^{-+}\gamma$, $0^{-+}\rightarrow 1^{+-} \gamma $\cite{vaishali21} for which very little data is available as of now, besides transitions such as: $0^{++}\rightarrow 1^{--} \gamma$ and $1^{--}\rightarrow 0^{++} \gamma$\cite{bhatnagar20}. We have made use of the generalized method\cite{bhatnagar20,vaishali21} of handling such quark triangle diagrams  in the framework of $4\times 4$ BSE.

In present calculation, we have been able to express $M_{fi}$ as a linear superposition of terms (similar to \cite{bhatnagar20, vaishali21}) involving combinations of $++$, and $--$ components of Salpeter wave functions of final hadron, with the terms involving $\Psi^{++}$  being associated with a coefficients, $I_1'$ and $I_2'$, while the $\Psi^{--}$  being associated with $I_1''$ and $I_2'' $, which are the results of pole integration in the complex $\sigma$-plane ($\sigma=\frac{q.P}{P^2}$ being the component of internal momentum $q$ of initial hadron that is longitudinal to external momentum, $P$). This superposition of all possible terms is a feature of relativistic frameworks. Due to this after the pole integration over the fourth (longitudinal) component ($Md\sigma$)of internal hadron momentum, $q$ is carried out, the effective 3D form of amplitude is still relativistically covariant, as can be seen from the general structures of invariant amplitude, $M_{fi}$ for the processes $e^-e^+\rightarrow \gamma \chi_{c0,c1}$, and $e^-e^+\rightarrow \gamma\eta_c$, expressible in terms of their form factors, whose integrands for all the three processes are two-dimensional functions of both the internal hadron momentum, $\hat{q}$, as well as the center-of-mass energy, $\sqrt{s}$. The 3D plots of these integrands reveal discontinuous regions for range of variables, $0< \hat{q} < 3$ GeV, and $4 < \sqrt{s} < 12$ GeV. The cross section for any process at a given $\sqrt{s}$ is calculated after removal of these discontinuities from $d^3\hat{q}$ integration involved in form factor calculations. Plots of cross sections for all these processes for the range of energies,  $4<\sqrt{s} <12$ GeV are drawn to explain the cross sections in both high ($\sqrt{s}$=10 - 12 GeV), and low ($\sqrt{s}$= 4 - 6 GeV) energy regions. These plots for all the three processes show mild variations in the low energy region as also observed in BESIII\cite{ablikim21,ablikim17} energy region, and are analyzed in terms of the form factors.

The paper is organised as follows: Section 2 deals with the BS equation for $Q\bar{Q}$ system, Section 3 deals with calculation of cross section for the process, $e^- e^+ \rightarrow \gamma \chi_{c0}$, while section 4 deals with the corresponding calculation for the process, $e^-  e^+ \rightarrow \gamma  \chi_{c1}$. The calculation of cross section for $e^-e^+\rightarrow \gamma \eta_c$ is dealt in Section 5. Section 6 deals with the Discussions.

\section{Bethe-Salpeter equation for $Q\bar{Q}$ bound state}
The Bethe-Salpeter equation (BSE) for a meson (a $q\bar{q}$ bound state) can be written as,
\begin{equation}
S_{F}^{-1}(p_{1})\Psi(P,q)S_{F}^{-1}(-p_{2}) =
i\int \frac{d^{4}q'}{(2\pi)^{4}}K(q,q')\Psi(P,q'),
\end{equation}

with $p_1$, $p_2$ being the momenta of the two particles of massses $m_1$ and $m_2$ respectively. $P$ is the external momentum of the hadron of mass, $M$, while $q$ is its internal momentum. In Eq.(1), $K(q,q')$ is the interaction kernel, and $S_{F}^{-1}(\pm p_{1,2})=\pm i{\not}p_{1,2}+ m_{1,2}$ are the inverse propagators for the quark and antiquark.

To reduce the above equation to 3D form, we make use of the Covariant Instantaneous Ansatz \cite{mitra01} on the BS kernel, $K(q,q')$, where we can write
$K(q,q')=K(\widehat{q},\widehat{q}')$, where the BS kernel depends entirely on the component of
internal momentum of the hadron, $\widehat{q}_\mu= q_\mu- \frac{q.P}{P^2}P_\mu$, which is a 3D variable, and is orthogonal to the total
hadron momentum, i.e. $\widehat{q}.P=0$, while $\sigma
P_\mu=\frac{q.P}{P^2}P_\mu$ is the component of $q$ that is longitudinal
to $P$. And the 4-dimensional volume element is,
$d^4q=d^3\widehat{q}Md\sigma$.

The 3D BS wave function, $\psi(\hat{q}')$ is obtained by integrating the 4D BS wave function, $\Psi(P,q')$ over the longitudinal component, $Md\sigma'$ of the four dimensional volume element, $d^4q'$ as,

\begin{equation}
\psi(\hat{q}')=\frac{i}{2\pi}\int Md\sigma' \Psi(P,q'),
\end{equation}

Now, it is to be observed that the longitudinal component $M\sigma'$  of $q'$ does not appear in $K(\hat{q},\hat{q}')$. We thus carry out integration over the longitudinal component $Md\sigma$ of the four-dimensional volume element $d^4q$ on right hand side of Eq.(1), and making use of the previous equation, we get,

\begin{equation}
S_F^{-1}(p_1)\Psi(P,q)S_F^{-1}(p_2)=\int \frac{d^3\hat{q}}{(2\pi)^3}K(\widehat{q},\widehat{q}')\psi(\hat{q}')=\Gamma(\hat{q}).
\end{equation}

Thus, we can identify the 4D hadron-quark vertex function, $\Gamma(\hat{q})$, which is the amputated 3-point function, and is directly related to the 4D BS wave function, $\Psi(P,q)$, which is a 3-point function. $\Psi(P,q)$ is obtained by sandwiching the hadron-quark vertex function, $\Gamma(\hat{q})$ between the two quark propagators as (see \cite{eshete19} and references therein),

\begin{eqnarray}
&&\nonumber \Psi(P, q)=S_1(p_1)\Gamma(\hat q)S_2(-p_2),\\&&
 \Gamma(\hat q)=\int\frac{d^3\hat q'}{(2\pi)^3}K(\hat q,\hat q')\psi(\hat q').
\end{eqnarray}

We wish to point out that the 4D BS wave function, $\Psi(P,q)$ is analogous to the quark bilinear 3-point correlation function used in a recent lattice calculation \cite{sternbeck19} of nonperturbative structure of vector and axial vector vertices. Here $\Gamma(\hat{q})$  is used for calculation of transition amplitudes of various processes.

A series of steps then leads to four Salpeter equations, which are the effective 3D forms of BSE \cite{eshete19}:

\begin{eqnarray}
 &&\nonumber(M-\omega_1-\omega_2)\psi^{++}(\hat{q})=\Lambda_{1}^{+}(\hat{q})\Gamma(\hat{q})\Lambda_{2}^{+}(\hat{q})\\&&
   \nonumber(M+\omega_1+\omega_2)\psi^{--}(\hat{q})=-\Lambda_{1}^{-}(\hat{q})\Gamma(\hat{q})\Lambda_{2}^{-}(\hat{q})\\&&
\nonumber \psi^{+-}(\hat{q})=0.\\&&
 \psi^{-+}(\hat{q})=0\label{fw5}
\end{eqnarray}

where, $\omega_{1,2}^{2}=m_{1,2}^2+\hat{q}^2$, and $\Lambda_{i}^{\pm}=\frac{1}{2\omega_i}[\frac{{\not}P}{M}\omega_{i}'\pm I(i)(im_i+\hat{q})]$ (where i=1,2, and $I(i)=(-1)^{i+1}$) are  the projection operators\cite{bhatnagar18,wang16,wang06} corresponding to constituents $\#$ 1 and $\#$ 2, in a hadron of rest mass, $M$, and external momentum, $P$, which project out the $\pm\pm$ components from the hadronic wave function, $\psi(\hat{q})$ as, $\psi^{\pm\pm}(\hat{q})=\Lambda_1^{\pm}(\hat{q})\frac{{\not}P}{M}\psi(\hat{q})\frac{{\not}P}{M}\Lambda_2^{\pm}(\hat{q})$. The sum of  projection operators satisfies the relation, $\Lambda_{i}^+ +\Lambda_{i}^-=\frac{{\not}P}{M}$. And the relation $\frac{{\not}P}{M}\frac{{\not}P}{M}=-1$, leads to $\frac{{\not}P}{M}(\Lambda_{i}^{+}+\Lambda_{i}^{-})=-1$.

Also, $\Gamma(\hat{q})$ in the first two Salpeter equations is the non-perturbative 4D hadron-quark vertex function (see Eqs.{3-4]) that enters into the 4D BS wave function, $\Psi(P,q)$. We start with a $4\times 4$ BSE which is fully covariant in all its details, after which we perform a 3D reduction on BSE under Covariant Instantaneous Ansatz- a Lorentz-invariant generalization of Instantaneous Approximation.

The 3D Salpeter equations in Eq.(5), have explicit dependence on the variable, $\hat{q}^2$, whose most important property is its positive definiteness $\hat{q}^2=q^2-\frac{(q.P)^2}{P^2}\geq 0$ on the hadron mass shell $P^2=-M^2$ throughout the entire 4D space\cite{eshete19, bhatnagar20 ,vaishali21}, and is a Lorentz-invariant variable\cite{bhatnagar20, wang16}, and a scalar, whose validity extends over the entire 4D space, while keeping contact with the surface $P.q=0$ (hadron rest frame), where $\hat{q}^2=\vec{q}^2$.  Further, it can be checked that the component, $\hat{q}_{\mu}$, is always orthogonal to the external hadron momentum, $P_{\mu}$ and satisfies the relation, $\hat{q}.P=0$, irrespective of whether $q.P=0$ (i.e. $\sigma=0$), or $q.P\neq 0$ (i.e. $\sigma \neq 0$). This condition $P.q=0$ is in fact the same as instantaneous approximation\cite{bhatnagar20, wang16}.

Further, it is to be noted that besides the 3D Salpeter equations, that are used for mass spectral predictions, the hadron-quark vertex function, $\Gamma(\hat{q})$, which is a  part of the 4D BS wave function as in Eq.(4), also has an explicit dependence on the variable, $\hat{q}^2$, due to which $\Gamma(\hat{q})$\cite{eshete19,bhatnagar20, vaishali21} enjoys an unrestricted access over the entire time-like region of the 4D space since $\hat{q}^2\geq 0$ for time-like hadron momentum, and is used for evaluating transition amplitudes through quark-loop diagrams over a range of energies.

For processes calculated here, the 3D form of transition amplitudes $M_{fi}$ obtained after pole integration over the longitudinal component $Md\sigma$ of the 4D volume element, $d^4 q$, has an explicit dependence on $\hat{q}^2$, and is again relativistically invariant, and thus frame independent for all the three processes studied in this work. This can be checked from Eqs.(25), (40) and (53) for $M_{fi}$, that are expressed in terms of the respective form factors.

Thus, in view of these remarkable properties of $\hat{q}_{\mu}$, which makes it an effectively 3D vector, our both the objectives: (i) 3D structure of BSE as the controlling equation for spectra, and (ii) a general enough (off-shell) structure of BS vertex function $\Gamma(\hat{q})$ to facilitate applications to transition amplitudes in 4D form is largely met if the BS kernel depends on $\hat{q}_{\mu}$. Further, in this approach, an important aspect is the appearance of the hadron-quark vertex $\Gamma(\hat{q})$ (used for calculation of transition amplitudes) on the right side of effectively 3D Salpeter equations (used for calculation of spectra) in Eq. (5), which gives a formal dynamical link between low energy mass spectroscopy and high energy transition amplitudes. The dynamical links between 3D spectra and 4D transition amplitudes have been shown in \cite{bhatnagar14} by showing the exact interconnection between the 3D and 4D BSE. Due to all the above features, the framework of $4\times 4$ BSE under Covariant instantaneous ansatz, is not only fully relativistic, but this approach also goes well beyond potential models. Further, the 3D reduction of BSE using the above covariant instantaneous ansatz is being used in recent works \cite{wang22, wang16, he21,yu19,vaishali21, chang10,zgwang05}.

Also in our work, Eqs.(14-15) are the expressions of quark propagators, expressed in terms of the wave function projection operators, $\Lambda^{+-}$. However we use constituent quark mass in these propagators, though in principle, the dressed quark propagators arise as the solution of the gap equation\cite{bhagwat06,cdroberts11,cdroberts11a}, which is characterized by a momentum-dependent mass function, $m(p)$.  It has been shown that constituent quark masses in the propagators of heavy (c,b) quarks is a good approximation \cite{zgwang05,chang10}, and provides a rationale for constituent quark masses employed for heavy quarks in potential models, and the constituent quarks masses in the propagators for heavy (c,b) quarks have also been employed in recent BSE calculations in \cite{wang22,yu19,wang16,he21,shi21,vaishali21,chang10,zgwang05}. Thus in the present work, we will use constituent quark masses for c,b quarks in propagators. However for future work, we do intend to use dressed quark propagators, for which the model will have to be modified.

Our work is based on QCD motivated BSE in ladder approximation, which is an approximate description\footnote{Such effective forms of the BS kernel
in ladder BSE have recently been used in \cite{wang22,he21,karmanov19,fredrico14,karmanov17,wang19}, and can predict bound states having a purely relativistic origin \cite{karmanov19}}, with an effective four-fermion interaction
mediated by a gluonic propagator\cite{eshete19}, that serves as the kernel of BSE in the lowest order. We can generalize this to any arbitrary interaction, $K$, where $K$ can be said to represent the sum of all irreducible graphs. The precise form of our kernel includes a confining term along with a one-gluon exchange term. Thus, the interaction kernel in BSE is taken to be vector type to make connection with the QCD degrees of freedom. It is one-gluon-exchange like as regards the colour and spin dependence, and thus has a general structure \cite{eshete19},

\begin{eqnarray}
&&\nonumber K(\hat{q},\hat{q}')=(\frac{1}{2}\lambda_1.\frac{1}{2}\lambda_2)\gamma_{\mu}\times \gamma_{\mu} V(\hat{q}, \hat{q}')\\&&
\nonumber V(\hat{q},\hat{q}')= \frac{4\pi\alpha_s(M^2)}{(\hat{q}-\hat{q}')^2}
 +\frac{3}{4}\omega^2_{q\bar q}\int d^3r\bigg(\kappa r^2-\frac{C_0}{\omega_0^2}\bigg)e^{i(\hat q-\hat q').\vec r}\equiv V_{OGE}(\hat{q},\hat{q}')+V_{conf.}(\hat{q},\hat{q}'),\\&&
\nonumber \omega_{q\bar{q}}^2=M\omega_0^2\alpha_s(M^2),\\&&
\alpha_s(M^2)=\frac{12\pi}{(33-2f)}[Log(M^2/\Lambda^2)]^{-1}.
\end{eqnarray}

The use of one-gluon-exchange kernel in our BSE framework preserves the connection with the gauge invariance. The scalar part of the kernel, $V(\hat{q},\hat{q}')\equiv V(\hat{k})$, with $\hat{k}_{\mu}=\hat{q}_{\mu}-\hat{q}_{\mu}'$, and $\hat{k}_{\mu}=k_{\mu}-\frac{k.P}{P^2}P_{\mu}$ is transverse to $P_{\mu}$, and $\hat{k}^2 \geq 0$ and is a four-scalar over the entire 4-dimensional space. Here $V(\hat{k})$ is written as a sum of the perturbative (one-gluon-exchange) part and the non-perturbative (confinement) part as in previous equation, while $ \kappa=(1+4\hat m_1\hat m_2A_0M^2r^2)^{-\frac{1}{2}}$.  The presence of running coupling constant, $\alpha_s$ in flavour dependent spring constant, $\omega_{q\bar{q}}^2$ provides an explicit QCD motivation to the BSE kernel. To do numerical calculations, we need to fix the scale in strong coupling constant, $\alpha_s(Q^2)$. We have thus fixed this scale as the meson mass, M. Thus we use $\alpha_s(M^2)$ in the interaction kernel on lines of \cite{eshete19,wang16}.

It is seen that the algebraic form of the confining potential ensures a smooth transition from nearly harmonic (for $c\bar{u}$) to almost linear (for $b\bar{b}$)as is believed to be true for QCD (see\cite{eshete19} for details regarding the nature of the confining potential).

Regarding the parameters of the model, $\omega_0$ = 0.22 GeV. is the spring constant, $A_0=0.01$, and $C_0$ = 0.69 are dimensionless constants, while $\frac{C_0}{\omega_0^2}$
plays the role of ground state energy, $\Lambda$ = 0.250 GeV is the QCD length scale, with input heavy quark masses, $m_c$ = 1.490GeV , and $m_b$ = 4.690 GeV. Our previous studies on mass spectral
calculations of heavy-light quarkonia \cite{bhatnagar20,eshete19} were used to fit the input parameters of our model.

We further wish to mention, that with the above form of confinement part of kernel, we have made successful predictions of mass spectra of ground and excited states of $0^{++}, 0^{-+}, 1^{--}, 1^{+-}$, and $1^{++}$ heavy-light quarkonia\cite{hluf16,eshete19,bhatnagar18,vaishali21a}, and the radial wave functions $\phi(\hat{q})$ obtained as solutions their mass spectral equations, have been used to calculate their leptonic decays\cite{eshete19,vaishali21a}, two-photon decays\cite{hluf16,bhatnagar18}, and their radiative $M1$ and $E1$ decays\cite{bhatnagar20,vaishali21} using the same set of input parameters that were fixed from the mass spectrum. We are further employing this framework for calculations of production cross sections of $e^- e^+\rightarrow \gamma \chi_{c0,c1}/\eta_c$ in this work.

Further, the important aspect of our BSE framework is that hadron-quark vertex $\Gamma(\hat{q})$ is derived from 4D BSE for both heavy and light mesons. Also, as regards the 4D BS wave functions, $\Psi(P,q)$ of $0^{++}, 1^{++}$, and $0^{-+}$ mesons in our present work are concerned, we start with their most general forms as in \cite{smith69,alkofer02}, which are consistent with the canonically normalized 4D BS wave functions in \cite{bhagwat06}, with the full Dirac structures of these BS wave functions of all the three hadrons employed, to maintain their completeness in this work.

\section{Cross section for $e^- + e^+ \rightarrow \gamma +\chi_{c0}$}
We consider both the $s$-channel diagrams in Fig.1 for the process, $e^- +e^+ \rightarrow \gamma^*\rightarrow \gamma +\chi_{c0,c1}$ , and calculate the contributions of both these diagrams to cross section. Here, $e^- e^+$ of momenta $\bar{p}_1$ and $\bar{p}_2$ annihilate to produce a virtual photon of momentum, $k'=\bar{p}_1+\bar{p}_2$, whose coupling to $c\bar{c}$ meson and photon is through the quark loop diagram as in Fig.1. These diagrams involve three electromagnetic vertices, and one non-perturbative strong vertex, which involve parity, $P$ and charge conjugation, $C$ conservation, and are hence governed by Landau-Yang theorem. Here, the photon-quark-anti-quark vertex is given as $ie_Q\gamma_{\mu}$, where, $e_Q=\frac{2}{3}e$ is the charge of the $c$ quark.

\begin{figure}[h!]
 \centering
 \includegraphics[width=15cm,height=4cm]{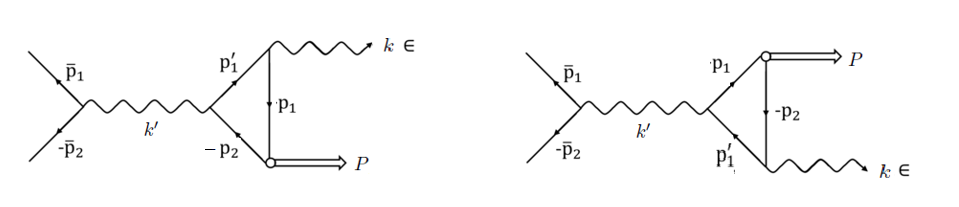}
 \caption{Lowest order $s$-channel Feynman diagrams for the production of $\gamma$ and $\chi_{cJ}(J=0,1)$ in electron-positron annihilation. The exchange diagram on the right is obtained from the diagram on the left by reversing the direction of internal fermionic lines.}
\end{figure}

For the diagrams in Fig.1, the invariant amplitude $M^{1}_{fi}$ for the first diagram on the left is given by the
one-loop momentum integral as:
\begin{equation}
M^1_{fi}=-ie e_Q^2 [\bar{v}^{(s2)}({\bar{p}}_2)\gamma_{\mu}u^{(s1)}(\bar{p}_1)]\frac{-1}{s}\int \frac{d^4q}{(2\pi)^4}Tr[\bar{\Psi}_S(P,q){\not}\epsilon^{\lambda'} S_F(p_1')\gamma_{\mu}],
\end{equation}

where, $P, q$ are the external momentum and internal momentum of $\chi_{c0}$, with 4D BS wave function, $\Psi_S(P,q)$, while, $k$ and $\epsilon^{\lambda'}$, are the momentum and the polarization vector of the emitted photon, and in the center of mass frame, we have expressed $\bar{p}_1+\bar{p}_2=\sqrt{s}$. The second diagram is obtained from the first diagram by reversing the direction of internal fermion lines, which amounts to exchange of final states. The amplitude from the second diagram is equal to the amplitude from the first diagram.  The wave functions of both incoming and outgoing particles are normalized to one particle per unit volume.

We start with the most general form of 4D BS wave function for scalar meson $0^{++}$, that is expressed in terms of various Dirac structures in \cite{smith69,alkofer02}. Then, making use of the 3D reduction under Covariant Instantaneous Ansatz, and making use of the fact that $\hat{q}.P=0$, we can write the general decomposition of the instantaneous BS wave function for scalar mesons of dimensionality $M$ being composed of various Dirac structures that are multiplied with scalar functions $f_i(\hat{q})$, and various powers of the meson mass $M$ as \cite{bhatnagar18,hluf16,wang22},

\begin{equation}
\Psi_S(\hat{q})=Mf_1(\hat{q})-i{\not}Pf_2(\hat{q})-i{\not}\hat{q}f_3({\not}\hat{q})-\frac{2{\not}P{\not}\hat{q}}{M}f_4(\hat{q}).
\end{equation}

These amplitudes $f_1,...f_4$ in the equation above are all independent. We put Eq.(8) into the Salpeter equations, Eq.(5). The last two Salpeter equations provide constraint relations between the amplitudes, where $f_1$ is expressed in terms of $f_3$, while $f_2$ is expressed in terms of $f_4$, and thus leading to the wave function, $\psi(\hat{q})$\cite{eshete19} being expressed in terms of $f_3$ and $f_4$ respectively. Then, putting this wave function into the first two Salpeter equations leads to two coupled integral equations in $f_3$ and $f_4$ with the structure of the interaction kernel incorporated. These equations are then  decoupled and reduced to two identical algebraic equations in amplitudes $f_3$ and $f_4$, leading to $f_3 \approx f_4$. We thus obtain the relativistic 3D wave function\cite{eshete19},

\begin{eqnarray}
&&\nonumber \Psi_S(\hat{q})=N_S\bigg[\frac{(m_1+m_2)\hat{q}^2}{\omega_1\omega_2+m_1m_2-\hat{q}^2}+i{\not}\hat{q}+2i\frac{(\omega_1-\omega_2)\hat{q}^2{\not}P}{M(\omega_1\omega_2+m_1m_2)}+\frac{2{\not}P{\not}\hat{q}}{M}\bigg]\phi_S(\hat{q})\\&&
\omega_{1,2}^2=m_{1,2}^2+\hat{q}^2,
\end{eqnarray}

which for equal mass scalar quarkonia such as $\chi_{c0}$ is reduced to:
\begin{equation}
\Psi_S(\hat{q})=N_S\bigg[\frac{\hat{q}^2}{m}+i{\not}\hat{q}+2\frac{{\not}P{\not}q}{M}\bigg]\phi_S(\hat{q})
\end{equation}

The mass spectrum is obtained by analytically solving the 3D Salpeter equations in Eq.(5). Solutions of these spectral equations in an approximate harmonic oscillator basis, lead to both the mass spectrum, as well as the algebraic forms of 3D radial meson wave functions, $\phi_S(\hat{q})$ (see \cite{bhatnagar18,eshete19,vaishali21a} for details):

\begin{eqnarray}
&&\nonumber \phi_S(1P,\hat q)=\sqrt{\frac{2}{3}}\frac{1}{\pi^{3/4}}\frac{1}{\beta_S^{5/2}} |\hat q| e^{-\frac{\hat q^2}{2\beta_S^2}},\\&&
 \phi_S(2P,\hat q)=\sqrt{\frac{5}{3}}\frac{1}{\pi^{3/4}}\frac{1}{\beta_S^{5/2}}|\hat q|(1-\frac{2\hat q^2}{5\beta_S^2})e^{-\frac{\hat q^2}{2\beta_S^2}},
\end{eqnarray}

where, $|\hat{q}|=\sqrt{q^2 – (q.p)^2/P^2}$, and is the length of the 3D vector $\hat{q}$, and is also a Lorentz-invariant variable (please see \cite{bhatnagar20}), and $\beta_S=(\frac{m\omega^2_{q\bar{q}}}{\sqrt{1+2A_0(N+3/2)}})^\frac{1}{4}$ \cite{eshete19} is the inverse range parameter \cite{bhatnagar18,eshete19}.

Now, the 3D meson wave functions, $\phi(\hat{q})$\cite{eshete19,vaishali21a} in Eqs.(11,36,49) are obtained as analytic solutions of mass spectral equations in an approximate harmonic oscillator basis \cite{eshete19, bhatnagar20, vaishali21a}, which are in turn obtained by decoupling the 3D Salpeter equations, and have been used to calculate various processes such as leptonic decays\cite{eshete19}, two-photon decays\cite{bhatnagar18,hluf16}, and radiative M1 and E1 decays \cite{bhatnagar20,vaishali21}for ground and excited states of heavy-light quarkonia. In the present work we have employed these wave functions for calculation of production processes of quarkonia in the framework of Bethe-Salpeter equation on lines of \cite{shi21}.

The 4D adjoint BS wave functions of $\chi_{c0}$ meson can be written as,
\begin{equation}
\bar{\Psi}_S(P, q)=S_F(-p_2)\bar{\Gamma}_S(\hat{q})S_F(p_1),
\end{equation}

Here, we consider $p_{1,2}$ as the momenta of quark/anti-quark of the scalar meson, $\chi_{c0}$ of total momentum, $P$. These quark momenta can be expressed in terms of its total and the internal momentum  as,
\begin{equation}
p_{1,2}=\frac{1}{2}P \pm q.
\end{equation}

We study this process in the center of mass frame where photon and $\chi_{c0}$ are emitted back to back, with $\overrightarrow{k}=-\overrightarrow{P}$. Considering the photon momentum, $k=(\overrightarrow{k}, i|\overrightarrow{k}|)$, it can be checked the $k.\hat{q}=0$, where $\hat{q}=q-\frac{q.P}{P^2}P$ is the component of internal momentum of the hadron, that is transverse to its external momentum, i.e. $P.\hat{q}=0$.

Since photon is transversely polarized, $\epsilon'=(\overrightarrow{\epsilon}',i0)$, we can express $P.\epsilon'$ as, $P.\epsilon'=\overrightarrow{P}.\overrightarrow{\epsilon}'=-\overrightarrow{k}.\overrightarrow{\epsilon}'$. Further, the orthogonality of $k$ with $\epsilon'$ implies,   $k.\epsilon'=\overrightarrow{k}.\overrightarrow{\epsilon}'=0$, and it can be checked that $P.\epsilon'=0$.

We now express the two quark propagators in $\bar{\Psi}_S(P, q)$ in terms of the projection operators, $\Lambda^+$ and $\Lambda^-$ as \cite{bhatnagar18,bhatnagar20,vaishali21},
\begin{eqnarray}
&&\nonumber S_F(p_1)=\frac{\Lambda^+_1(\hat{q})}{\eta_1}+\frac{\Lambda^-_1(\hat{q})}{\eta_2}\\&&
\nonumber S_F(-p_2)=\frac{-\Lambda^+_2(\hat{q})}{\eta_3}+\frac{-\Lambda^-_2(\hat{q})}{\eta_4};\\&&
\nonumber \eta_{1,2}=M\sigma+\frac{1}{2}M\mp \omega_1;\\&&
\eta_{3,4}=-M\sigma+\frac{1}{2}M\mp \omega_2,
\end{eqnarray}

where $\eta_1,...\eta_4$ are the denominators of various terms in the quark propagators that are expressed in terms of the longitudinal component, $M\sigma$ of the internal hadron momentum, $q$, that is to be integrated over, while $\omega_{1,2}^2=m_{1,2}^2+\hat{q}^2$. Due to the fact that the momentum of the third quark, $p_1'=k+p_1$, where $p_1=\frac{1}{2}P+\hat{q}+\sigma P$ (where we have decomposed the internal momentum of the hadron as $q=\hat{q}+\sigma P$), we can express the propagator, $S_F(p_1')$ as
\begin{eqnarray}
&&\nonumber S_F(p'_1)=\frac{-i(\not k+\frac{1}{2}\not P+\not \hat{q}+\sigma \not P)+m}{-M^2\sigma^2-(4E^2+M^2)\sigma-\alpha}\\&&
\alpha=\hat{q}^2+m^2-2E^2-\frac{M^2}{4},
\end{eqnarray}

where, $E=\sqrt{P^2+M^2}$ is the energy of the final meson. We now try to reduce $M^1_{fi}$ in Eq.(7) to 3D form. For this, we put the above expressions for quark propagators in $M^1_{fi}$, making use of the fact that $q=(\hat{q},M\sigma)$, and the four-dimensional volume element,
$d^4q= d^3\hat{q} Md\sigma$, we split the 4D volume integral in $M^1_{fi}$ as,

\begin{eqnarray}
&&\nonumber M^1_{fi}=-ie e_Q^2 [\bar{v}^{(s2)}({\bar{p}}_2)\gamma_{\mu}u^{(s1)}(\bar{p}_1)]\frac{-1}{s}\times\\&&
\int \frac{d^3\hat{q}}{(2\pi)^3}\int \frac{Md\sigma}{2\pi}Tr\bigg[\bigg((\frac{-\Lambda^+_2(\hat{q})}{\eta_3}+\frac{-\Lambda^-_2(\hat{q})}{\eta_4})
\bar{\Gamma}_S(\hat{q})(\frac{\Lambda^+_1(\hat{q})}{\eta_1}+\frac{\Lambda^-_1(\hat{q})}{\eta_2})\bigg)\not\epsilon' S_F(p'_1)\gamma_{\mu}\bigg],
\end{eqnarray}

where the quantity in trace bracket can in turn be expressed by means of Salpeter equations in Eq.(5) as,

\begin{eqnarray}
&&\nonumber M^1_{fi}=-ie e_Q^2 [\bar{v}^{(s2)}({\bar{p}}_2)\gamma_{\mu}u^{(s1)}(\bar{p}_1)]\frac{-1}{s}\times\\&&
\int \frac{d^3\hat{q}}{(2\pi)^3}\int \frac{Md\sigma}{2\pi i}Tr\bigg[\bigg(-\frac{\bar{\Psi}^{++}(\hat{q})(M-2\omega)}{\eta_3\eta_1}-\frac{\bar{\Psi}^{--}(\hat{q})(M+2\omega)}{\eta_4\eta_2}\bigg)\not\epsilon' S_F(p'_1)\gamma_{\mu}\bigg],
\end{eqnarray}
where the $+-$ and $-+$ terms vanish due to the last two Salpeter equations, $\bar{\Psi}^{+-}(\hat{q})=\bar{\Psi}^{-+}(\hat{q})=0$ in Eq.(5) (that have acted as the constraint equations in derivation of mass spectrum). Due to equal mass mesons considered in this work, we take, $\omega_1=\omega_2=\omega=\sqrt{m^2+\hat{q}^2}$.

The above equation after insertion of $\eta_1,...\eta_4$, and $S_F(p'_1)$ from Eq.(14-15) is reexpressed as,
\begin{eqnarray}
&&\nonumber M^1_{fi}=-ie e_Q^2[\bar{v}^{(s2)}({\bar{p}}_2)\gamma_{\mu}u^{(s1)}(\bar{p}_1)]\frac{-1}{s}(\frac{-1}{M^2})(\frac{-1}{M^2})\times\\&&
\nonumber \int \frac{d^3\hat{q}}{(2\pi)^3}\int \frac{Md\sigma}{2\pi i}
Tr\bigg[\bigg(-\frac{\bar{\Psi}^{++}(\hat{q})(M-2\omega)}{[\sigma-(-\frac{1}{2}+\frac{\omega}{M})][\sigma-(\frac{1}{2}-\frac{\omega}{M})]}-\frac{\bar{\Psi}^{--}(\hat{q})(M+2\omega)}{[\sigma-(-\frac{1}{2}-\frac{\omega}{M})][\sigma-(\frac{1}{2}+\frac{\omega}{M})]}\bigg)\times \\&&
\not\epsilon' \frac{-i(\not k+\frac{1}{2}\not P+\not \hat{q}+\sigma \not P)+m}{[\sigma-(-\frac{1}{2}-\frac{2E^2}{M^2}+\frac{1}{M}\sqrt{\omega^2+\frac{4E^4}{M^2}})][\sigma-(-\frac{1}{2}-\frac{2E^2}{M^2}-\frac{1}{M}\sqrt{\omega^2+\frac{4E^4}{M^2}})]}\gamma_{\mu}\bigg].
\end{eqnarray}

We now carry out $Md\sigma$ integration over the poles of the propagators, $S_F(\pm p_{1,2})$ and $S_F(p'_1)$ in complex $\sigma$-plane, with pole positions:
\begin{eqnarray}
&&\nonumber \sigma_1^{\pm}=-\frac{1}{2}\mp \frac{\omega}{M}\pm i\epsilon\\&&
\nonumber \sigma_2^{\pm}=\frac{1}{2}\mp \frac{\omega}{M}\pm i\epsilon\\&&
\beta^{\pm}=(-\frac{1}{2}-\frac{2E^2}{M^2})\mp \frac{1}{M}\sqrt{\omega^2+\frac{4E^4}{M^2}}\pm i\epsilon.
\end{eqnarray}

It can be checked that the results of integration, whether one closes the contour above or below the $\sigma$-plane is the same. After performing integration over $Md\sigma$, we can express $M^1_{fi}$ as,

\begin{eqnarray}
&&\nonumber M^1_{fi}=-ie e_Q^2[\bar{v}^{(s2)}({\bar{p}}_2)\gamma_{\mu}u^{(s1)}(\bar{p}_1)]\frac{-1}{s}(\frac{-1}{M^2})(\frac{-1}{M^2})\times\\&&
\nonumber \int \frac{d^3\hat{q}}{(2\pi)^3}
Tr\bigg[[\bar{\Psi}_S^{++}(\hat{q})(M-2\omega)I'_1+\bar{\Psi}_S^{--}(\hat{q})(M+2\omega)I''_1] [\not \epsilon' (-i(\not k+\frac{1}{2}\not P+\not \hat{q})\gamma_{\mu})+\not \epsilon' m\gamma_{\mu}]+\\&&
~~~~~~[\bar{\Psi}_S^{++}(\hat{q})(M-2\omega)I'_2+\bar{\Psi}_S^{--}(\hat{q})(M+2\omega)I''_2](-i)\not \epsilon' \not P\gamma_{\mu}\bigg],
\end{eqnarray}

where, $I_1',I_1'',I_2'$ and $I_2''$ in previous equation, are the results of $Md\sigma$ integrations in the complex $\sigma$-plane over the poles of propagators, $S_F(\pm p_{1,2})$, and $S_F(p_1')$ in Fig.1 and are given in Eq.(57) in Appendix.

Regarding the nature of these integrals, it can be checked that the  expressions for  $I_{1,2}^{‘}$, and $I_{1,2}^{''}$   are quite general, where the only input parameter present in them is the quark mass, $m$ that enters into the expression for relativistic frequency, $\omega=\sqrt{m^2 +\hat{q}^2}$. The only other quantities present are: the mass, $M$ of the produced hadron and the center of mass energy, $\sqrt{s}=2E$. Thus, $I'_{1,2}$, and $I''_{1,2}$ are two-dimensional functions of internal hadron momentum,  $\hat{q}$ and center of mass energy, $\sqrt{s}$. To check their continuity, we did  3D plots of these four integrals versus ($\hat{q}$, $\sqrt{s}$) along two axes in Fig.2. Discontinuities in the region, $0 < \hat{q} <3 $ for all $4<\sqrt{s}<12$ are observed in the integrals, $I'_1$, and $I'_2$ that are characterized by break in the function, though the integrals, $I_{1}''$, and $I_{2}''$ are continuous over the entire range of values of $\hat{q}$, and for $4 < \sqrt{s} < 12$ GeV, and form a continuous sheet. These integrals, $I_{1,2}^{',''}$ will be used in evaluation of transition amplitudes, $M_{fi}$ for various processes.

\begin{figure}[h!]
 \centering
 \includegraphics[width=15cm,height=8cm]{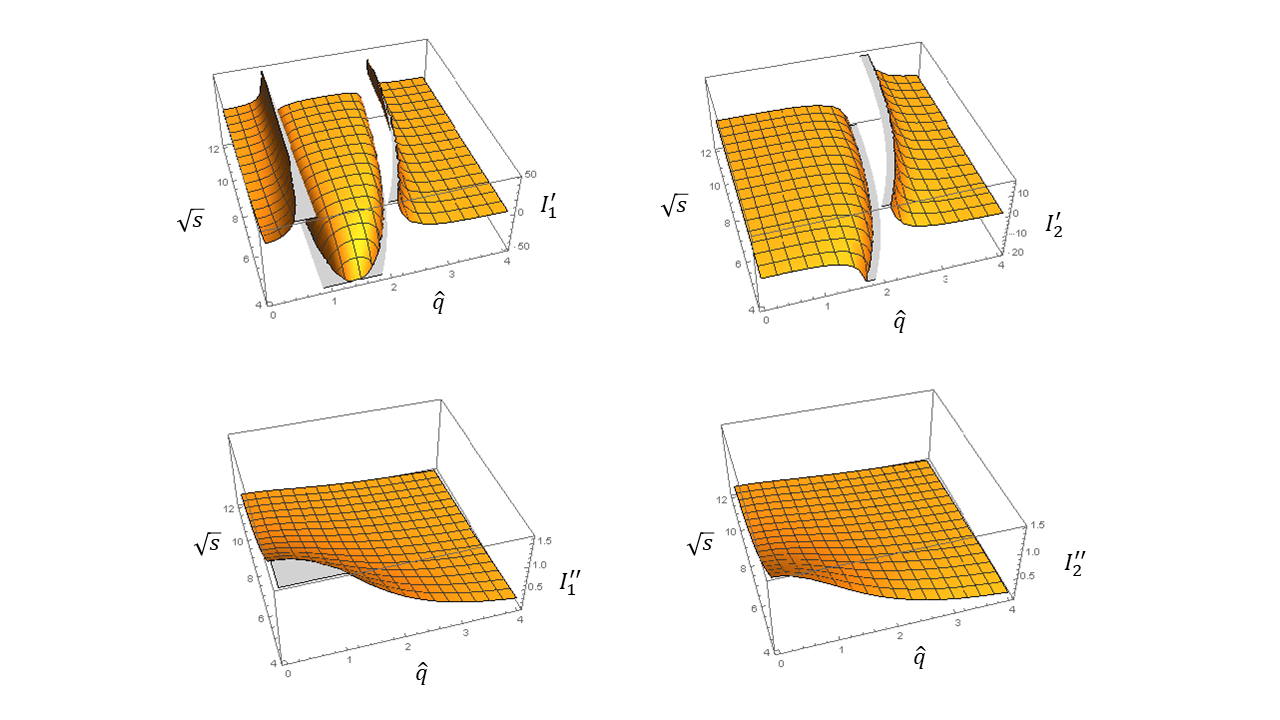}
 \caption{3D plots of the countour integrals $I_1' ,I_2'$ (in first line) and $I_1'', I_2''$ (in second line) as functions of variables, $\hat{q}$ and $\sqrt{s}$ for the range, $0 < \hat{q} < 4 $ GeV, and $4 < \sqrt{s} < 6$ GeV.}
\end{figure}

Now, lets combine together the $\bar{\Psi}^{++}$ and $\bar{\Psi}^{--}$ terms in Eq.(20). For this, we first write the adjoint wave functions, $\bar{\Psi}^{++}(\hat{q})$ and $\bar{\Psi}^{--}(\hat{q})$ in the general form:
\begin{eqnarray}
&&\nonumber \bar{\Psi}^{++}(\hat{q})=N_S\phi_s(\hat{q})[a_1+ib_1\not P+ic_1 \not \hat{q} +d_1 \not \hat{q} \not \hat P]\\&&
\nonumber \bar{\Psi}^{--}(\hat{q})=N_S\phi_s(\hat{q})[a_2+ib_2\not P+ic_2 \not \hat{q} +d_2 \not \hat{q} \not \hat P]\\&&
\nonumber a_1=a_2=\frac{\hat{q}^2}{4m}+\frac{m\hat{q}^2}{4\omega^2}+\frac{\hat{q}^4}{4\omega^2 M}-\frac{\hat{q}^2}{\omega}\\&&
\nonumber b_{1}=\frac{\hat{q}^2}{2\omega M}-\frac{m\hat{q}^2}{M\omega^2}\\&&
\nonumber c_{1}= \frac{1}{4}+\frac{m^2}{4\omega^2}+\frac{\hat{q}^2}{2\omega^2}-\frac{m}{\omega}\\&&
\nonumber d_{1}=\frac{\hat{q}^2}{2\omega m M}+\frac{m}{2\omega M}+\frac{1}{2M}+\frac{m^2}{2\omega^2 M}+\frac{\hat{q}^2}{2\omega^2 M}\\&&
\nonumber b_{2}=\frac{\hat{q}^2}{2\omega M}+\frac{m\hat{q}^2}{M\omega^2}\\&&
\nonumber c_{2}= \frac{1}{4}+\frac{m^2}{4\omega^2}-\frac{\hat{q}^2}{2\omega^2}-\frac{m}{\omega}\\&&
d_{2}=-\frac{\hat{q}^2}{2\omega m M}+\frac{m}{2\omega M}+\frac{1}{2M}+\frac{m^2}{2\omega^2 M}-\frac{\hat{q}^2}{2\omega^2 M},
\end{eqnarray}
where, $a_{1,2},b_{1,2},...$ are the coefficients associated with the Dirac structures, $1, {\not}P,{\not}q,...$ etc, $m$ being the quark mass, and $N_S$ being the 4D BS amplitude normalization, that is obtained through the current conservation condition, and the algebraic expressions for the 3D BS wave functions $\phi_S(\hat{q})$ for both ground and excited states of scalar mesons are given in Eq.(11). We can then express $M^1_{fi}$ as,
\begin{eqnarray}
&&\nonumber M^1_{fi}=\frac{iee_Q^2}{M^4}[\bar{v}^{(s2)}({\bar{p}}_2)\gamma_{\mu}u^{(s1)}(\bar{p}_1)]\frac{-1}{s}N_S\int \frac{d^3 \hat{q}}{(2\pi)^3}\phi_S(\hat{q}) \times \\&&
\nonumber Tr\bigg[(X_1+i\not P X_2 +i\not \hat{q} X_3+i\not \hat{q} \not P X_4)(-i)\not \epsilon' (\not k+\frac{1}{2}\not P+\not \hat{q})\gamma_{\mu}+
(Y_1+i\not P Y_2 +i\not \hat{q} Y_3+i\not \hat{q} \not P Y_4)(-i)\not \epsilon' \not P \gamma_{\mu}+\\&&
~~~~~~~~~~~~~~~~~~~~~~~~~~~~~~~~~~~~~~(X_1+i\not P X_2 +i\not \hat{q} X_3+i\not \hat{q} \not P X_4)\not \epsilon' \gamma_{\mu} m\bigg],
\end{eqnarray}

where,

\begin{eqnarray}
&&\nonumber X_1=a_1(M-2\omega)I'_{1}+a_2(M+2\omega)I''_{1}\\&&
\nonumber Y_1=a_1(M-2\omega)I'_{2}+a_2(M+2\omega)I''_{2}\\&&
\nonumber X_2=b_1(M-2\omega)I'_{1}+b_2(M+2\omega)I''_{1}\\&&
\nonumber Y_2=b_1(M-2\omega)I'_{2}+b_2(M+2\omega)I''_{2}\\&&
\nonumber X_3=c_1(M-2\omega)I'_{1}+c_2(M+2\omega)I''_{1}\\&&
\nonumber Y_3=c_1(M-2\omega)I'_{2}+c_2(M+2\omega)I''_{2}\\&&
\nonumber X_4=d_1(M-2\omega)I'_{1}+d_2(M+2\omega)I''_{1},\\&&
Y_4=d_1(M-2\omega)I'_{2}+d_2(M+2\omega)I''_{2}.
\end{eqnarray}

Evaluating trace over the gamma matrices, we can express $M^1_{fi}$ in Eq.(22) as,

\begin{eqnarray}
&&\nonumber
M^1_{fi}=4i\frac{ee_Q^2}{M^4}[\bar{v}^{(s2)}({\bar{p}}_2)\gamma_{\mu}u^{(s1)}(\bar{p}_1)]\frac{-1}{s}N_S\int \frac{d^3\hat{q}}{(2\pi)^3}\phi_S(\hat{q})
[(\alpha_1+\alpha_4 \hat{q}^2)\epsilon'_{\mu}+\alpha_2(\hat{q}.\epsilon') P_{\mu}+\alpha_3(\hat{q}.\epsilon') k_{\mu}];\\&&
\nonumber \alpha_1=-4X_2 (P.k)+2X_2 M^2 -4X_3 \hat{q}^2 +4M^2Y_2+4X_1 m;\\&&
\nonumber \alpha_2=(2X_3 +4X_2+4Y_3-4X_4 m)\\&&
\nonumber \alpha_3=4X_3,\\&&
\alpha_4=8X_3,
\end{eqnarray}
where expressions for $X_1,...,X_4, Y_1,...,Y_4$ are given in Eqs.(23). All expressions for $\alpha_1,...\alpha_4$ involve the quark mass, $m$.

Now, since $\hat{q}$ is an effective 3D variable, we can write $\hat{q}.\epsilon'=\hat{q}_{\nu}\epsilon'_{\nu}=|\hat{q}|(I.\epsilon')$, where, $I_{\mu}$ is a unit vector along the direction of $\hat{q}_{\nu}$, and is expressed as, $I=\frac{\hat{q}}{|\hat{q}|}$, where $|\hat{q}|=\sqrt{\hat{q}^2}=\sqrt{q^2-(q.P)^2/P^2}$, and is a Lorentz-invariant variable \cite{bhatnagar20,wang06,wang16}.

\bigskip

Total amplitude for the process, $e^- + e^+ \rightarrow \gamma +\chi_{c0}$ will be the sum of contributions from both the diagrams of Fig.1, i.e. $M_{fi}=M^I_{fi} + M^2_{fi}$. Since the contribution from both the diagrams is equal, the total amplitude for process is two times the amplitude for any of the process, i.e. $M_{fi}=2M^1_{fi}$. Thus, we can express $M_{fi}$ as,
\begin{eqnarray}
&&\nonumber M_{fi}=i[\bar{v}^{(s2)}({\bar{p}}_2)\gamma_{\mu}u^{(s1)}(\bar{p}_1)]
\bigg[\beta_1\epsilon'_{\mu}+\beta_2(I.\epsilon') P_{\mu}+\beta_3(I.\epsilon') k_{\mu}\bigg];\\&&
\nonumber \beta_1=\frac{8ee_Q^2 N_S}{M^4 s}
\int \frac{d^3\hat{q}}{(2\pi)^3}\phi_S(\hat{q})(\alpha_1+\alpha_4 \hat{q}^2);\\&&
\nonumber \beta_2=\frac{8ee_Q^2 N_S}{M^4 s}\int \frac{d^3\hat{q}}{(2\pi)^3}\phi_S(\hat{q})|\hat{q}|\alpha_2;\\&&
\beta_3=\frac{8ee_Q^2 N_S}{M^4 s}\int \frac{d^3\hat{q}}{(2\pi)^3}\phi_S(\hat{q})|\hat{q}|\alpha_3,
\end{eqnarray}

where, $\beta_1,...,\beta_3$ are the form factors, which are functions of the center-of-mass energy, $\sqrt{s}$. The above expression for $M_{fi}$ can be expressed as,
\begin{eqnarray}
&&\nonumber M_{fi}=i[\bar{v}^{(s2)}({\bar{p}}_2)\gamma_{\mu}u^{(s1)}(\bar{p}_1)]M_{\mu};\\&&
M_{\mu}=\beta_1\epsilon'_{\mu}+\beta_2(I.\epsilon') P_{\mu}+\beta_3(I.\epsilon') k_{\mu},
\end{eqnarray}

where $M_{\mu}$ is the amplitude for the transition, $\gamma*\rightarrow \gamma \chi_{c0}$, which is the matrix element of electromagnetic current, $J_{\mu}$ between the vacuum and the $|\gamma,\chi_{c0}>$ state (expressed as, $<\gamma,\chi_{c0}|J_{\mu}|0>$). Now, electromagnetic gauge invariance demands that, $k_{\mu}M_{\mu}=0$. This condition leads to $\beta_2=0$, due to free photon in final state satisfying Einstein condition, and with its polarization vector, $\epsilon'$ being transverse to its momentum, $k$. Thus we are left with two independent form factors, $\beta_1$ and $\beta_3$. And we obtain $M_{fi}$ as,

\begin{equation}
M_{fi}=[\bar{v}^{s_2}(p_2)\gamma_{\mu}u^{s_1}(p_1)][\beta_1\epsilon'_{\mu}+\beta_3(I.\epsilon') k_{\mu}],
\end{equation}

Thus, this transition amplitude simplifies to a general form required by Lorentz covariance as shown in the previous equation. The expressions for the $\beta_1$ and $\beta_3$ in Eq.(25), involve integrals over $d^3\hat{q}$, and absorb all the momentum dependence of $M_{fi}$, each of which have an explicit dependence on the charmed quark mass, $m$ (through the variables, $\alpha_1,...,\alpha_4$, and $X_1/Y_1,...,X_4/Y_4$), whose numerical values are listed in Table 3.

Thus, the  transition amplitude, $M_{fi}$ in Eq.(27) involves form factors $ \beta_1$ and $\beta_3$, with their expressions in Eq.(25), where their integrands involve $\alpha_i$, which through Eqs.(23-24) are in turn expressed as super positions of the contour integrals, $I_1',I_1'',I_2'$ and $I_2''$ (where $I_1’$ and $I_2’$ have discontinuities as shown by their 3D plots in Fig.2) , which are multiplied by factors $(M-2\omega)$, and $(M+2\omega)$, that are again functions of $\hat{q}$.  Therefore, it would be logical to study the continuity structure of the integrands, $I_{\beta_1}(\hat{q},\sqrt{s})$, and $I_{\beta_3}(\hat{q},\sqrt{s})$ of the form factors, $\beta_1$ and $\beta_3$ respectively, in variables, $\hat{q}$ and $\sqrt{s}$, for which we do 3D plots of their integrands for range of variables, $0< \hat{q} <4$ GeV, and $4 < \sqrt{s} < 12$ GeV in Fig.3.

\begin{figure}[h!]
 \centering
 \includegraphics[width=15cm,height=6cm]{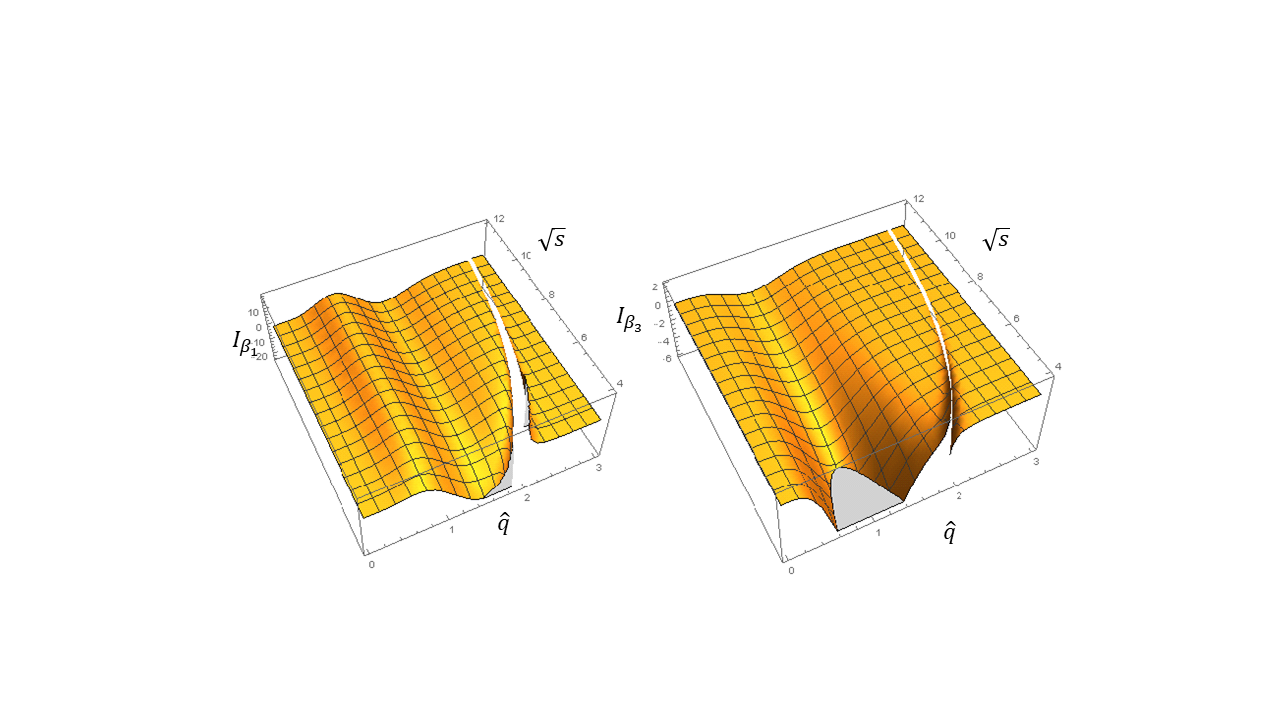}
 \caption{3D plots of integrands $I_{\beta_1}$, and $I_{\beta_3}$ of form factors, $\beta_1$, and $\beta_3$ respectively, as functions of $\hat{q}$ and $\sqrt{s}$ for range of variables, $0< \hat{q} <3$ GeV, and $4 < \sqrt{s} < 12$ GeV for the process, $e^- e^+ \rightarrow \gamma \chi_{c0}$.}
\end{figure}

From these 3D plots of integrands of $\beta_1$ and $\beta_3$, one can see that the discontinuities in both these graphs lie in a narrow region 1.5 GeV $<\hat{q}<$3 GeV for the range of energies, 4 GeV$ < \sqrt{s} <$ 12 GeV. Also, it is seen that as one goes from 4 GeV to 12GeV, the region of discontinuity keeps  getting narrowed, and keeps shifting to the right on the $\hat{q}$-axis, as is apparent in  Fig.4. Thus to evaluate the form factors $\beta_1$ and $\beta_3$, (and in turn, the cross sections) for any value of the center of mass energy, $\sqrt{s}$, the region of discontinuity has to be precisely determined, and removed from these integrals. To do this, we do 2D plots of integrands of form factors, $\beta_1$ and $\beta_3$ versus the internal hadron momentum, $\hat{q}$ for each center of mass energy, $\sqrt{s}$. To illustrate this procedure, we give these plots of integrands at three center of mass energies $\sqrt{s}$=4 GeV, 8 GeV, and 12 GeV, for the process, $e^- e^+ \rightarrow \gamma \chi_{c0}$. in Fig.4.

\begin{figure}[h!]
 \centering
 \includegraphics[width=15cm,height=12cm]{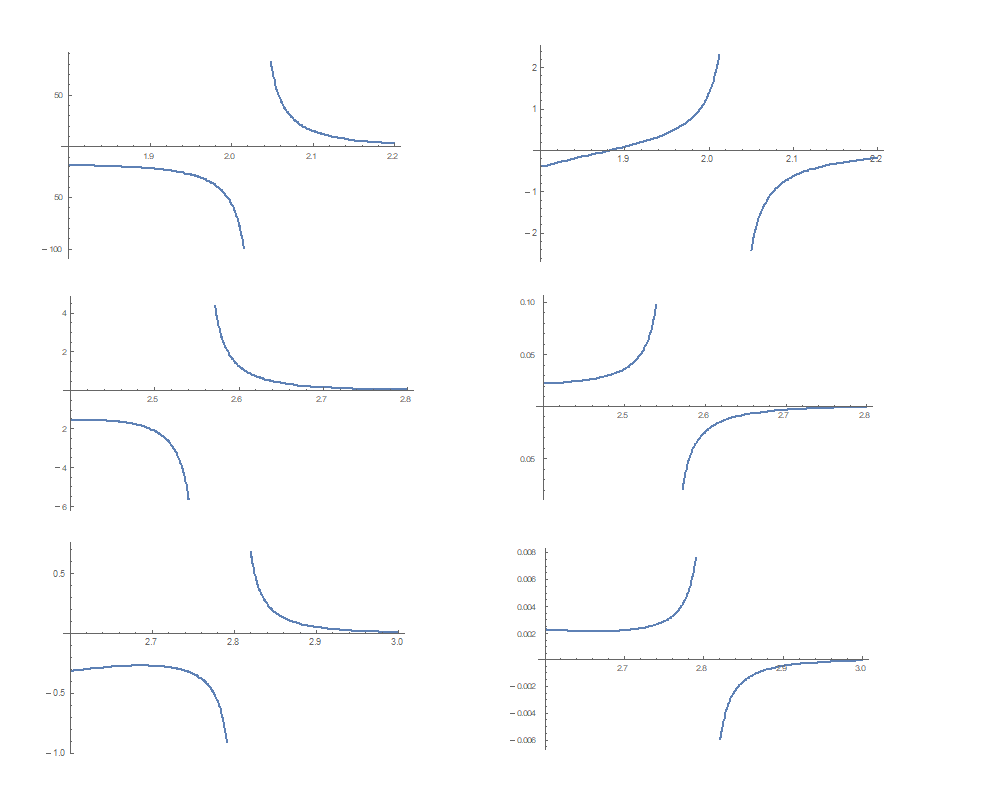}
 \caption{Two-dimensional plots of integrands $I_{\beta_1}$, and $I_{\beta_3}$ of form factors, $\beta_1$, and $\beta_3$ respectively, as functions of $\hat{q}$ at three different center of mass energies $\sqrt{s}$=4.6 GeV (in line 1), 8 GeV (in line 2) and 12 GeV (in line 3) to precisely identify the region of discontinuity in $d^3\hat{q}$ integration in Eq.(25), for the process, $e^- e^+ \rightarrow \gamma \chi_{c0}$ at a given center of mass energy.}
\end{figure}

The region of discontinuity up to two decimal places is thus identified, and removed from the form factor integrals, $\beta_1$ and $\beta_3$ for each value of $\sqrt{s}$ at which we intend to calculate the cross section.  The regions of discontinuity for some values of $\sqrt{s}$ for the process, $e^- e^+\rightarrow \gamma\chi_{c0}$ are listed in Table 1, and a similar procedure is followed to identify and remove the region of discontinuity at any other $\sqrt{s}$.

\begin{table}[htp]
  \begin{center}
  \begin{tabular}{p{2cm}p{4cm}p{4cm}}
  \hline
 $\sqrt{s}$ &  $\hat{q}$ (for $\beta_1$) & $\hat{q}$ (for $\beta_3$)\\
  \hline
  12  &2.79$<\hat{q}<$ 2.82&2.79$<\hat{q}<$ 2.82\\
  10.6&2.73$<\hat{q}<$2.76&2.73$<\hat{q}<$2.76   \\
  8    &2.54$<\hat{q}<$ 2.57& 2.54$<\hat{q}<$ 2.57\\
  6    &2.28 $<\hat{q}<$ 2.32   &2.27 $<\hat{q}<$ 2.33 \\
  4.6    &2.01$<\hat{q}<$2.05   &2.01$<\hat{q}<$ 2.05   \\
       \hline
  \end{tabular}
\caption{Regions of discontinuities in hadron internal momentum, $\hat{q}$(in GeV) in the integrands $I_{\beta_1}$, and $I_{\beta_3}$ of form factors $\beta_1$, and $\beta_3$ respectively, in Eq.(25) at different center of mass energies, $\sqrt{s}$ (in GeV) in the range, $4.6 - 12$ GeV for the process $e^- e^+ \rightarrow \gamma \chi_{c0}(1P)$.}
\end{center}
\end{table}

The spin averaged amplitude modulus square,

\begin{equation}
|\bar{M}_{fi}|^2 = \frac{1}{4}\sum_{s_1, s_2, \lambda} M^{\dag}_{fi} M_{fi}
\end{equation}
is obtained by averaging over the spins of incident particles, and summing over the polarizations of the final photon.

\bigskip

Thus we obtain
\begin{equation}
|\bar{M}_{fi}|^2=\bigg[4\beta_1^2(\bar{p}_1.\bar{p}_2+m_e^2)+8\beta_3^2(\bar{p}_1.k)(\bar{p}_2.k)\bigg].
\end{equation}

We take the electromagnetic coupling constant, $\alpha_{em}=\frac{e^2}{4\pi}$, while, $m_e$ and $m$ being the masses of electron and $c$-quark respectively. We study this process in the center of mass frame, where scalar meson and photon are emitted back to back, i.e. $\overrightarrow{P}=-\overrightarrow{k}$, with ($|\overrightarrow{P}|=|\overrightarrow{k}|$), and similarly, $\overrightarrow{p}_2=-\overrightarrow{p}_1$ (where $\bar{p}_1$ and $\bar{p}_2$ are the momenta of incoming electron and positron respectively). Considering $\theta$ to be angle between $\overrightarrow{p}_2$ and $\overrightarrow{P}$, we express, $\bar{p}_1+\bar{p}_2=\sqrt{s}$, ~$\bar{p}_1.\bar{p}_2=\frac{1}{2}(-s+2m_e^2)$, where $\sqrt{s}=2E=E_{cm}$, being the center of mass energy. Similarly the dot products of momenta can be expressed as: $\bar{p}_1.P=\bar{p}_2.k \approx E^2(1+cos\theta)$,  $\bar{p}_1.k=\bar{p}_2.P\approx -E^2(1-cos\theta)$, and $P.k=\frac{1}{2}(-s+2M^2)$, where $M$ is the mass of the final scalar meson. Further, the square of virtual photon momentum, $k'^2=(\bar{p}_1+\bar{p}_2)^2=s$.

We can then express $|\bar{M}_{fi}|^2$ as,

\begin{equation}
|\bar{M}_{fi}|^2=\bigg[2\beta_1^2(-s+3m_e^2)-\frac{1}{2}\beta_3^2s^2(1-cos^2\theta)]\bigg]
\end{equation}

The cross terms associated with the product, $\beta_1\beta_3$ are zero on account of trace calculations. Further, $N_S$ is the 4D BS amplitude normalization of scalar meson, and is obtained through the current conservation condition,

\begin{equation}\label{46}
2iP_\mu=\int \frac{d^{4}q}{(2\pi)^{4}}
\mbox{Tr}\left\{\overline{\Psi}(P,q)\left[\frac{\partial}{\partial
P_\mu}S_{F}^{-1}(p_1)\right]\Psi(P,q)S_{F}^{-1}(-p_2)\right\} +
(1\rightleftharpoons2).
\end{equation}

Following a series of steps, we can express the total cross section for the process as,

\begin{equation}
\sigma=\frac{1}{32\pi^2 s^{3/2}}|\vec{P'}|\int d\Omega' |\bar{M}_{fi}|^2,
\end{equation}

with $|\vec{P'}|=\frac{s-M^2}{\sqrt{s}}$ being the momenta of either of the outgoing particles in the center of mass frame, and $|\bar{M}_{fi}|^2$, being expressed as in Eq.(30). As regards the dimensionalities of various quantities, it can be checked that the BS normalizer for $\chi_{c0}$, $N_S \sim M^{-3}$, and the form factors, $\beta_1\sim M^{-1}$, and $\beta_3\sim M^{-2}$. Thus the cross sectional formula behaves as, $\sigma \sim M^{-2}$.

Thus we have expressed the amplitude and the cross section for the process, $e^+ e^-\rightarrow \gamma \chi_{c0}$ in terms of the radial wave functions of scalar quarkonia, which were analytically obtained from solutions\cite{eshete19} of mass spectral equations of scalar mesons. The cross sections for $e^- e^+\rightarrow \gamma \chi_{c0}(nP)$ calculated in this work are given in Table 1 along with results of other models.

\begin{table}[htp]
  \begin{center}
  \begin{tabular}{p{3.4cm}p{0.8cm}p{1.8cm}p{2.9cm}p{1.5cm}p{1.3cm}p{0.8cm}p{1.3cm}p{0.8cm}}
  \hline
 Process&$\sqrt{s}$ &BSE & Experiment&\cite{braguta10}& \cite{brambilla20} &\cite{sang20}& \cite{chung21}&\cite{li13}\\
  \hline
   $e^- e^+\rightarrow \gamma \chi_{c0}(1P)$&10.6&3.810   &$<205.9$\cite{jia18} &6.1$\pm$ 3.9  & $1.84^{+0.25}_{-0.26}$ &2.52 & $3.11^{+0.05}_{-0.94}$&1.81\\
   $e^- e^+\rightarrow \gamma \chi_{c0}(1P)$&4.6&46.617   &$<2.6\times 10^{3}$\cite{ablikim21} &  &   & &&131\\
   $e^- e^+\rightarrow \gamma \chi_{c0}(1P)$&4.0&13.785   &$<4.5\times 10^{3}$\cite{ablikim21} &  &   & &&1173\\
  $e^- e^+\rightarrow \gamma \chi_{c0}(2P)$&10.6&3.570   & &             &           &&&1.04\\
  $e^- e^+\rightarrow \gamma \chi_{c0}(2P)$&4.6&64.073   & &  & &   & &962\\
  \hline
  \end{tabular}
\caption{Cross sections for processes, $e^- e^+\rightarrow \gamma \chi_{c0} (nP)~ (n=1,2)$ (in fb) calculated in BSE-CIA at $\sqrt{s}=10.6 GeV$, 4.6 GeV and 4 GeV, along with recent data from Belle\cite{jia18}(at 10.58 GeV), BESIII\cite{ablikim21} (at 4.6 GeV and 4 GeV), and other models.}
\end{center}
\end{table}

The numerical values of $N_S, \beta_1$, and $\beta_3$ for process $e^- e^+\rightarrow \gamma \chi_{c0}$, calculated in our work at $\sqrt{s}=10.6GeV.$, and $4.6 GeV.$, are listed in Table 3 below.

\begin{table}[hhhh]
  \begin{center}
  \begin{tabular}{p{3.5cm} p{1.5cm} p{1.5cm} p{2.9cm} p{2.9cm} }
  \hline
 Process&                                 $N_S$&$\sqrt{s}$ & $\beta_1$  & $\beta_3$ \\
  \hline
  $e^- e^+\rightarrow \gamma \chi_{c0}(1P)$&5.630&10.6   &$-4.86966\times 10^{-6}$ & $-2.93112\times 10^{-6}$ \\
  $e^- e^+\rightarrow \gamma \chi_{c0}(2P)$&4.907&10.6   & $5.5833\times 10^{-6}$ & $2.967\times 10^{-6}$\\
  $e^- e^+\rightarrow \gamma \chi_{c0}(1P)$&5.630&4.6   &$-1.24247\times 10^{-4}$ & $-7.5177\times 10^{-5}$ \\
  $e^- e^+\rightarrow \gamma \chi_{c0}(2P)$&4.907&4.6   & $1.473\times 10^{-4}$ & $6.342\times 10^{-5}$\\
       \hline
  \end{tabular}
\caption{Numerical values of BS normalizer, $N_S$ (in $GeV^{-3}$) for $\chi_{c0}(1P)$, and $\chi_{c0}(3P)$, along with the numerical values of form factors, $\beta_1$ (in $GeV.^{-1}$), and $\beta_3$ (in $GeV^{-2}$) for processes, $e^- e^+\rightarrow \gamma \chi_{c0}(nP)~(n=1,2)$ at $\sqrt{s}=10.6$ GeV, and $4.6$ GeV }
\end{center}
\end{table}

\section{Cross section for $e^-+e^+\rightarrow \gamma+\chi_{c1}$}
We now study the process, $e^- +e^+ \rightarrow \gamma\rightarrow \gamma +\chi_{c1}$ , and calculate its cross section. These colour singlet leading-order (LO) Feynman diagrams for this process are given in Fig. 1. both of which contribute equally, and the total amplitude will be two times the amplitude from any one of the two diagrams.

The invariant amplitude $M^{1}_{fi}$ for $\gamma+\chi_{c1}$ production, corresponding to the first diagram on left in Fig.1, is given by the one-loop momentum integral as:
\begin{equation}
M^1_{fi}=-ie e_Q^2 [\bar{v}^{(s2)}({\bar{p}}_2)\gamma_{\mu}u^{(s1)}(\bar{p}_1)]\frac{-1}{s}\int \frac{d^4q}{(2\pi)^4}Tr[\bar{\Psi}_A(P,q){\not}\epsilon S_F(p_1')\gamma_{\mu}],
\end{equation}

where $P, q$ and $\epsilon^\lambda$ are the external momentum, internal momentum and the polarization vector of the axial meson, $\chi_{c1}$, while $k$ and $\epsilon^{\lambda'}$ are the momentum and the polarization vector of the emitted photon. We start with the 4D hadronic BS wave function, $\Psi_A(P,q)$ expressed in terms of various Dirac structures as in \cite{smith69,alkofer02}. The total amplitude for the process, $M_{fi}=2M^1_{fi}$.
After 3D reduction of the 4D BS wave function under Covariant Instantaneous Ansatz (CIA), the 3D BS wave function of dimension, $M$ can be expressed as\cite{bhatnagar18},

\begin{equation}
 \Psi_A(\hat{q})=\gamma_5{\not}\epsilon[iMf_1(\hat{q})+{\not}P f_2(\hat{q})-{\not}\hat{q} f_3(\hat{q}) +2i\frac{{\not}P{\not}\hat{q}}{M}f_4(\hat{q})]+\gamma_5 (\epsilon.\hat{q})[f_3(\hat{q})+2i\frac{{\not}P}{M} f_4(\hat{q})]
\end{equation}

Putting this wave function into the four Salpeter equations, and following a similar procedure as in case of scalar meson in previous section leads to $\Psi_A(\hat{q})$ being expressed as,
\begin{equation}
\Psi_A(\hat{q})=N_A \gamma_5\bigg[iM{\not}\epsilon +{\not}\epsilon{\not}P+2i\frac{{\not}\epsilon {\not}P {\not}\hat{q}}{M}\bigg]\phi_A(\hat{q})
\end{equation}

where $\phi_A(\hat{q})$ is the solution of the mass spectral equation\cite{vaishali21a} in approximate harmonic oscillator basis, obtained by analytically solving the four Salpeter equations, Eq.(5) for axial vector meson, and for ground and first excited states, their algebraic forms\cite{vaishali21a} are,

\begin{eqnarray}
&&\nonumber \phi_A(1P,\hat q)=\sqrt{\frac{2}{3}}\frac{1}{\pi^{3/4}}\frac{1}{\beta_A^{5/2}} |\hat q| e^{-\frac{\hat q^2}{2\beta_A^2}},\\&&
 \phi_A(2P,\hat q)=\sqrt{\frac{5}{3}}\frac{1}{\pi^{3/4}}\frac{1}{\beta_A^{5/2}}|\hat q|(1-\frac{2\hat q^2}{5\beta_A^2})e^{-\frac{\hat q^2}{2\beta_A^2}},
\end{eqnarray}

where $\beta_A=(\frac{2}{3}M\omega^2_{q\bar{q}})^{\frac{2}{3}}$\cite{vaishali21a} is the inverse range parameter. The adjoint BS wave function of $\chi_{c1}$ meson is given by,

\begin{equation}
\bar{\Psi}_A(P, q)=S_F(-p_2)\bar{\Gamma}_A(\hat{q})S_F(p_1),
\end{equation}

The expression for $M_{fi}$ in Eq.(33) involves integration over the four-dimensional volume element $d^4 q$, which is expressed as $d^4 q = d^3\hat{q} Md\sigma$. We again reduce this expression to 3D form by integrating over the longitudinal component, $Md\sigma$ over the poles of the propagators, $S_F(\pm p_{1,2})$ of quark and anti-quark, and the propagator for the third quark, $S_F(p_1')$, given as, $\sigma^{\pm}_{1,2}$, and $\beta^{\pm}$ in Eq.(19). The results of these contour integrations, $I'_1, I''_1, I'_2, I''_2$ are given in Eq.(56) of Appendix. The 3D plots of these integrals versus $\hat{q}$ and $\sqrt{s}$  are given in Fig.2, and their continuity discussed in Section 3. $M_{fi}$ is then expressed in the same form as Eq.(20), but with $\bar{\Psi}_A^{++}$ and $\bar{\Psi}_A^{--}$ in place of $\bar{\Psi}_S^{++}$ and $\bar{\Psi}_S^{--}$ as,

\begin{eqnarray}
&&\nonumber M_{fi}=-2ie e_Q^2[\bar{v}^{(s2)}({\bar{p}}_2)\gamma_{\mu}u^{(s1)}(\bar{p}_1)]\frac{-1}{s}(\frac{-1}{M^2})(\frac{-1}{M^2})\times\\&&
\nonumber \int \frac{d^3\hat{q}}{(2\pi)^3}
Tr\bigg[[\bar{\Psi}_A^{++}(\hat{q})(M-2\omega)I'_1+\bar{\Psi}_A^{--}(\hat{q})(M+2\omega)I''_1] [\not \epsilon' (-i(\not k+\frac{1}{2}\not P+\not \hat{q})\gamma_{\mu})+m\not \epsilon' \gamma_{\mu}]+\\&&
~~~~~~[\bar{\Psi}_A^{++}(\hat{q})(M-2\omega)I'_2+\bar{\Psi}_A^{--}(\hat{q})(M+2\omega)I''_2](-i)\not \epsilon' \not P\gamma_{\mu}\bigg].
\end{eqnarray}

The $++$ and $--$ components of $\bar{\Psi}_A(\hat{q})$ are:

\begin{eqnarray}
&&\nonumber \bar{\Psi}_A^{++}(\hat{q})=N_A\gamma_5[i{\not}\epsilon \theta_1+{\not}P{\not}\epsilon \theta_2+i{\not}\hat{q}{\not}\hat{P}{\not}\epsilon\theta_3+{\not}\hat{q}{\not}\epsilon\theta_4+i{\not}\hat{q}(\epsilon.\hat{q})\theta_5
+i{\not}P(\epsilon.\hat{q})\theta_6+(\hat{q}.\epsilon){\not}\hat{q}{\not}P\theta_7]\phi_A(\hat{q}),\\&&
\nonumber \bar{\Psi}_A^{--}(\hat{q})=N_A\gamma_5[i{\not}\epsilon \rho_1+{\not}P{\not}\epsilon \rho_2+i{\not}\hat{q}{\not}\hat{P}{\not}\epsilon\rho_3+{\not}\hat{q}{\not}\epsilon\rho_4+i{\not}\hat{q}(\epsilon.\hat{q})\rho_5
+i{\not}P(\epsilon.\hat{q})\rho_6+(\hat{q}.\epsilon){\not}\hat{q}{\not}P\rho_7]\phi_A(\hat{q}),\\&&
\nonumber \theta_1=\frac{M}{4}+\frac{mM}{2\omega}+\frac{m^2M}{4\omega^2}-\frac{M\hat{q}^2}{4\omega^2}+\frac{\hat{q}^2}{\omega},~ \theta_2=(\frac{1}{4}-\frac{m^2}{4\omega^2}+\frac{\hat{q}^2}{4\omega^2})\\&&
\nonumber \theta_3=\frac{1}{2M}+\frac{\hat{q}^2}{2M\omega^2},~ \theta_4=\frac{M}{2\omega}+\frac{m}{\omega},\\&&
\nonumber \theta_5=-\frac{1}{\omega},~ \theta_6=\frac{\hat{q}^2}{M\omega^2},~ \theta_7=\frac{m}{M\omega^2},\\&&
\nonumber \rho_1=\frac{M}{4}-\frac{mM}{2\omega}-\frac{m^2M}{4\omega^2}-\frac{\hat{q}^2}{\omega}-\frac{M\hat{q}^2}{4\omega^2}, ~ \rho_2=(-\frac{1}{4}+\frac{m^2}{4\omega^2}-\frac{\hat{q}^2}{4\omega^2})\\&&
\rho_3=\frac{1}{2M}-\frac{\hat{q}^2}{2\omega^2},~ \rho_4=-\frac{M}{2\omega}-\frac{m}{\omega},~ \rho_5=\frac{1}{\omega},~\rho_6=-\frac{\hat{q}^2}{\omega^2M},~\rho_7=\frac{m}{\omega^2M}.
\end{eqnarray}

Putting $\bar{\Psi}_A^{++}$, and $\bar{\Psi}_A^{--}$ from Eq.(39) into Eq.(38), defining a unit vector $I$ along the direction of $\hat{q}$ as in the previous section for $e^-e^+\rightarrow \gamma \chi_{c0}$, and carrying out trace over the gamma matrices, we can express the transition amplitude as,

\begin{eqnarray}
&&\nonumber M_{fi}=[\bar{v}^{s_2}(p_2)\gamma_{\mu}u^{s_1}(p_1)]\bigg[g_1\epsilon_{\mu\nu\alpha\beta}\epsilon^{\lambda}_{\nu}\epsilon^{\lambda'}_{\alpha}P_{\beta}+
g_2\epsilon_{\mu\nu\alpha\beta}\epsilon^{\lambda}_{\nu}\epsilon^{\lambda'}_{\alpha}k_{\beta}+
g_3(I.\epsilon) \epsilon_{\mu\nu\alpha\beta}P_{\nu}\epsilon^{\lambda'}_{\alpha}k_{\beta}\bigg]\\&&
\nonumber g_1=\frac{8ee_Q^2 N_A}{M^4s}\int \frac{d^3\hat{q}}{(2\pi)^3}\phi_A(\hat{q})(\frac{1}{2}\Theta_1+\Omega_1+\Theta_2m+\Theta_3\hat{q}^2),\\&&
\nonumber g_2=\frac{8ee_Q^2 N_A}{M^4s}\int \frac{d^3\hat{q}}{(2\pi)^3}\phi_A(\hat{q})\Theta_1,\\&&
g_3=\frac{8ee_Q^2 N_A}{M^4s}\int \frac{d^3\hat{q}}{(2\pi)^3}\phi_A(\hat{q})|\hat{q}|\Theta_6,
\end{eqnarray}

where, $g_1,...,g_3$ are the three form factors, with the quantities, $\Theta_1,...,\Theta_4$ and $\Omega_1,...,\Omega_4$ defined as,

\begin{eqnarray}
&&\nonumber \Theta_1=\theta_1(M-2\omega)I'_{1}+\rho_{1}(M+2\omega)I''_{1},\\&&
\nonumber \Omega_1=\theta_1(M-2\omega)I'_{2}+\rho_{1}(M+2\omega)I''_{2},\\&&
\nonumber \Theta_2=\theta_2(M-2\omega)I'_{1}+\rho_2(M+2\omega)I''_{1},\\&&
\nonumber \Theta_3=\theta_3(M-2\omega)I'_{1}+\rho_3(M+2\omega)I''_{1},\\&&
\nonumber \Theta_6=\theta_6(M-2\omega)I'_{1}+\rho_6(M+2\omega)I''_{1},\\&&
\end{eqnarray}

while $N_A$ is the BS amplitude normalization that is obtained from the current conservation condition, Eq.(31). From Eq.(40), we see that, $M_{fi}$ can again be expressed in a general form, $M_{fi}=[\bar{v}^{s_2}(p_2)\gamma_{\mu}u^{s_1}(p_1)]M_{\mu}$, with, $M_{\mu}=<\gamma,\chi_{c1}|J_{\mu}|0>$ being the transition matrix element of the electromagnetic current from $\gamma*\rightarrow \gamma\chi_{c1}$, which simplifies to a general form required by Lorentz-covariance, as,

\begin{equation}
M_{\mu}=[g_1\epsilon_{\mu\nu\alpha\beta}\epsilon^{\lambda}_{\nu}\epsilon^{\lambda'}_{\alpha}P_{\beta}+
g_2\epsilon_{\mu\nu\alpha\beta}\epsilon^{\lambda}_{\nu}\epsilon^{\lambda'}_{\alpha}k_{\beta}+
g_3(I.\epsilon) \epsilon_{\mu\nu\alpha\beta}P_{\nu}\epsilon^{\lambda'}_{\alpha}k_{\beta}],
\end{equation}

where, the form factors, $g_1, g_2$ and $g_3$  contain the entire momentum dependence of the amplitude $M_{fi}$. Now, the condition for gauge invariance, $k_{\mu}M_{\mu}=0$, leads to $g_1=0$. Thus, we can write $M_{fi}$ as,,

\begin{equation}
M_{fi}=[\bar{v}^{s_2}(p_2)\gamma_{\mu}u^{s_1}(p_1)][g_2\epsilon_{\mu\nu\alpha\beta}\epsilon^{\lambda}_{\nu}\epsilon^{\lambda'}_{\alpha}k_{\beta}+
g_3(I.\epsilon) \epsilon_{\mu\nu\alpha\beta}P_{\nu}\epsilon^{\lambda'}_{\alpha}k_{\beta}].
\end{equation}

It is to be noted that, the integrands, $I_{g_2}(\hat{q},\sqrt{s})$, and $I_{g_3}(\hat{q},\sqrt{s})$ of the form factors $g_2$ and $g_3$ in Eq.(40) are again functions of both $\hat{q}$ and $\sqrt{s}$, whose continuity has to be checked over the range of values of $\hat{q}$ and $\sqrt{s}$. For this, we again do a 3D plot of their integrands in Fig.5.

\begin{figure}[h!]
 \centering
 \includegraphics[width=15cm,height=6cm]{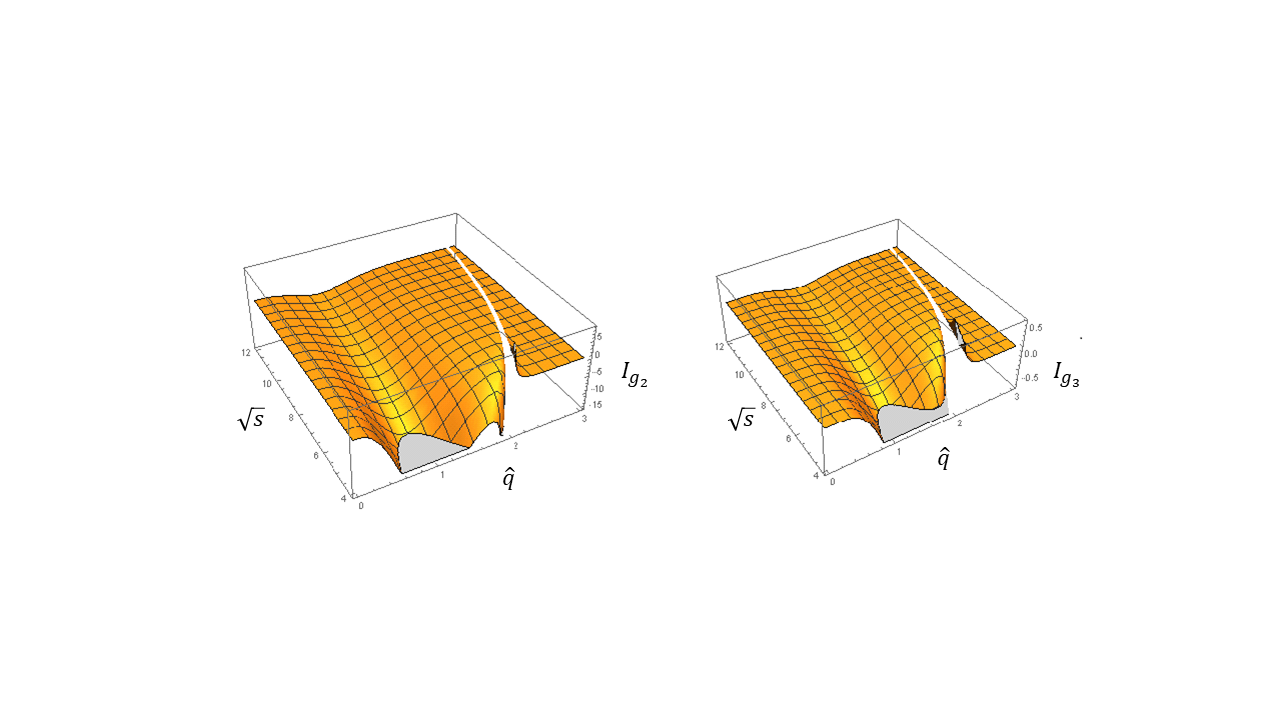}
 \caption{3D plots of integrands $I_{g_2}$, and $I_{g_3}$ of form factors, $g_2$, and $g_3$ respectively, as functions of $\hat{q}$ and $\sqrt{s}$ for range of variables, $0< \hat{q} <3$ GeV, and $4 < \sqrt{s} < 12$ GeV for the process, $e^- e^+ \rightarrow \gamma \chi_{c1}$.}
\end{figure}

The integrands of $g_2$ and $g_3$ are found to be discontinuous in the region $2<\hat{q} <3$ GeV, for $4< \sqrt{s} < 12$ GeV, and to precisely identify, and remove them from the integrals for any value of $\sqrt{s}$, we again do 2D plots of these integrands for a given $\sqrt{s}$ (as in Fig.4 for the $\gamma \chi_{c0}$ production process). We list the regions of discontinuity of these integrands for some values of $\sqrt{s}$ in Table 4.

\begin{table}[htp]
  \begin{center}
  \begin{tabular}{p{2cm}p{4cm}p{4cm}}
  \hline
 $\sqrt{s}$ &  $\hat{q}$ (for $g_2$) & $\hat{q}$ (for $g_3$)\\
  \hline
  12  &2.88$<\hat{q}<$ 2.91&2.88$<\hat{q}<$ 2.91\\
  10.6&2.82$<\hat{q}<$2.84&2.82$<\hat{q}<$2.84   \\
  8    &2.61$<\hat{q}<$ 2.65& 2.61$<\hat{q}<$ 2.65\\
  6    &2.35 $<\hat{q}<$ 2.39   &2.34 $<\hat{q}<$ 2.39 \\
  4.6    &2.07$<\hat{q}<$2.11   &2.07$<\hat{q}<$ 2.11   \\
       \hline
  \end{tabular}
\caption{Regions of discontinuities in hadron internal momentum, $\hat{q}$(in GeV) in the integrands $I_{g_2}$, and $I_{g_3}$ of form factors $g_2$, and $g_3$ respectively, in Eq.(25) at different center of mass energies, $\sqrt{s}$ (in GeV) in the energy range, ($4.6 - 12$) GeV for the process $e^- e^+ \rightarrow \gamma \chi_{c1}(1P)$.}
\end{center}
\end{table}

It can be again seen that the region of discontinuity keeps shifting to the right on the $\hat{q}$ axis, with increase in $\sqrt{s}$. The calculation of cross section, $\sigma$ involves evaluation of spin averaged amplitude modulus squared,

\begin{equation}
|\bar{M}_{fi}|^2= \frac{1}{4}\sum_{s_1,s_2,\lambda,\lambda'} |M_{fi}|^2,
\end{equation}

which can be expressed as,

\begin{eqnarray}
&&\nonumber |\bar{M}_{fi}|^2=\frac{1}{4}\bigg[\nonumber \frac{1}{48M^2}g_2^2[4(M^4+s^2)(-6m_e^2+s)+s^3(1+cos\theta)^2+M^2(48m_e^2s-8s^2-s^2(1+cos\theta)(-1+3cos\theta))]-\\&&
~~~~~~~~~~~~~~~~~~~~~~~~~~~~~\frac{1}{96}\frac{g_3^2}{3}(M^2-s)[64m_e^2s+s^2(5+3cos^2\theta)+M^2(-64m_e^2+6s(1+cos\theta))]\bigg].
\end{eqnarray}

The cross section can be calculated using Eqs.(32), with $|\overrightarrow{P}'|=\frac{s-M^2}{\sqrt{s}}$ being the momentum of final meson in the center of mass frame, defined in previous section. Since the BS Normalizer for $\chi_{c1}$, i.e. $N_A\sim M^{-3}$, while the form factors for the process, $e^- e^+ \rightarrow \gamma \chi_{c1}$ behave as,  $g_2\sim M^{-2}$, while $g_3\sim M^{-3}$, the cross section for the process again behaves as, $\sigma \sim M^{-2}$, whose numerical results calculated in our model are given in Table 5, along with experimental data and results of other models, while the numerical values of $N_A, g_2,g_3$ are given in Table 6.

\begin{table}[hhhhhhhh]
  \begin{center}
  \begin{tabular}{p{3.1cm}p{0.8cm}p{1cm}p{3.7cm}p{1.4cm}p{1.4cm}p{1.4cm}p{1.5cm}p{1.4cm} }
  \hline
 Process                              & $\sqrt{s}$&BSE&Experiment\cite{jia18,ablikim21} &\cite{brambilla20}& \cite{sang20}&\cite{chung21}&\cite{braguta10}&\cite{li13}\\
 $e^- e^+\rightarrow \gamma \chi_{c1}(1P)$&10.6&12.071&$17.3^{+4.2}_{-3.9}\pm1.7$ &$16.4^{+0.2}_{-0.2}$ &25.96&$23.3^{+7.9}_{-7.0}$&24.2$\pm$ 13 &9.7\\
 $e^- e^+\rightarrow \gamma \chi_{c1}(1P)$&4.6&75.832&$(1.7^{+0.8}_{-0.6}\pm0.2)\times 10^{3}$& & &  & & 1760\\
 $e^- e^+\rightarrow \gamma \chi_{c1}(1P)$&4.0&50.055&$(4.5^{+1.5}_{-1.3}\pm0.4)\times 10^{3}$& & &  & & 1173\\
  $e^- e^+\rightarrow \gamma \chi_{c1}(2P)$&10.6&9.751&   & & &  &  &20.8\\
   $e^- e^+\rightarrow \gamma \chi_{c1}(2P)$&4.6&43.518&   & & &  &  &4573\\
       \hline
  \end{tabular}
\caption{Cross sections for processes, $e^- e^+\rightarrow \gamma \chi_{c1} (nP)~ (n=1,2)$ (in fb) calculated in BSE-CIA at $\sqrt{s}$=10.6 GeV, 4.6 GeV, and 4 GeV along with recent data from Belle\cite{jia18}(at 10.58 GeV), BESIII\cite{ablikim21} (at 4.6 and 4 GeV), and other models}
\end{center}
\end{table}

\begin{table}[hhhh]
  \begin{center}
  \begin{tabular}{p{3.5cm} p{1.5cm} p{1.5cm} p{2.9cm} p{2.9cm} }
  \hline
 Process&                                 $N_A$&$\sqrt{s}$ & $g_1$  & $g_3$ \\
  \hline
  $e^- e^+\rightarrow \gamma \chi_{c1}(1P)$&5.4772&10.6   &$-6.00665\times 10^{-6}$ & $-3.45581\times 10^{-7}$ \\
  $e^- e^+\rightarrow \gamma \chi_{c1}(2P)$&4.9664&10.6   & $6.0455\times 10^{-6}$ & $6.8688\times 10^{-7}$\\
  $e^- e^+\rightarrow \gamma \chi_{c1}(1P)$&5.4772&4.6   &$-1.68999\times 10^{-4}$ & $-1.07590\times 10^{-5}$ \\
  $e^- e^+\rightarrow \gamma \chi_{c1}(2P)$&4.9664&4.6   & $1.865\times 10^{-4}$ & $2.194\times 10^{-5}$\\
       \hline
  \end{tabular}
\caption{Numerical values of BS normalizer, $N_A$ (in $GeV^{-3}$) for $\chi_{c1}(1P)$, and $\chi_{c1}(3P)$, along with the numerical values of $g_2$ (in $GeV.^{-2}$), and $g_3$ (in $GeV^{-3}$) for the processes, $e^- e^+\rightarrow \gamma \chi_{c1}(nP)~(n=1,2)$ at $\sqrt{s}=10.6 GeV$, and 4.6GeV}
\end{center}
\end{table}

\section{Cross section for $e^-+e^+\rightarrow \gamma+\eta_c$}
Again the two colour singlet leading-order (LO) Feynman diagrams for this process are given in Figures 1. both of which contribute equally, and the total amplitude will be two times the amplitude from any one of the two diagrams in Fig.1. The invariant amplitude $M^{1}_{fi}$ for $\gamma+\eta_c$ production, corresponding to the first diagram on left in Fig.1, is given by the one-loop momentum integral as:
\begin{equation}
M^1_{fi}=-ie e_Q^2 [\bar{v}^{(s2)}({\bar{p}}_2)\gamma_{\mu}u^{(s1)}(\bar{p}_1)]\frac{-1}{s}\int \frac{d^4q}{(2\pi)^4}Tr[\bar{\Psi}_P(P,q){\not}\epsilon^{\lambda'} S_F(p_1')\gamma_{\mu}],
\end{equation}

where $P, q$  are the external momentum and internal momentum of pseudoscalar meson, $\eta_c$ while $k$, and $\epsilon^{\lambda'}$ are the momentum and the polarization vector of the emitted photon.

As in the case of scalar ($0^{++}$), and axial vector ($1^{++}$) mesons in the previous sections, again, we start with the 4D hadronic BS wave function, $\Psi_P(P.q)$ expressed in terms of various Dirac structures as in \cite{smith69,alkofer02}. After 3D reduction of the 4D BS wave function under Covariant Instantaneous Ansatz (CIA), the 3D BS wave function of dimension, $M$ can be expressed as \cite{bhatnagar18,hluf16,wang22}},

\begin{equation}
\Psi_P(\hat{q})=N_P [M\phi_1(\hat{q})-i{\not}P\phi_2(\hat{q})+i{\not}\hat{q}\phi_3(\hat{q})+\frac{{\not}P{\not}\hat{q}}{M}\phi_4(\hat{q})]\gamma_5.
\end{equation}

Putting the above wave function into the Salpeter equations, Eq.(5), and following the same procedure as in the case of scalar mesons, leads to the relativistic 3D wave function for equal mass pseudoscalar mesons (for details see\cite{eshete19}):

\begin{equation}
\Psi_P(\hat{q})=N_P [M-i{\not}P+2\frac{{\not}P{\not}\hat{q}}{M}]\gamma_5\phi_P(\hat{q}).
\end{equation}

where $\phi_P(\hat{q})$ is the solution of the mass spectral equation\cite{hluf16,eshete19} in approximate harmonic oscillator basis, obtained by
analytically solving the four Salpeter equations, Eq.(5) for pseudoscalar meson, and for ground and first excited states, their
algebraic forms are:

\begin{eqnarray}
&&\nonumber \phi_P(1S,\hat q)=\frac{1}{\pi^{3/4}\beta_P^{3/2}} e^{-\frac{\hat q^2}{2\beta_P^2}},\\&&
 \phi_P(2S,\hat q)=\sqrt{\frac{3}{2}}\frac{1}{\pi^{3/4}\beta_P^{3/2}}(1-\frac{2\hat q^2}{3\beta_P^2}) e^{-\frac{\hat q^2}{2\beta_P^2}},
\end{eqnarray}

with $\beta_P$ being the inverse range parameter for pseudoscalar meson. We again express the amplitude for the first diagram as,

\begin{eqnarray}
&&\nonumber M^1_{fi}=-ie e_Q^2[\bar{v}^{(s2)}({\bar{p}}_2)\gamma_{\mu}u^{(s1)}(\bar{p}_1)]\frac{-1}{s}(\frac{-1}{M^2})(\frac{-1}{M^2})\times\\&&
\nonumber \int \frac{d^3\hat{q}}{(2\pi)^3}
Tr\bigg[[\bar{\Psi}_P^{++}(\hat{q})(M-2\omega)I'_1+\bar{\Psi}_P^{--}(\hat{q})(M+2\omega)I''_1] [\not \epsilon' (-i(\not k+\frac{1}{2}\not P+\not \hat{q})\gamma_{\mu})+m\not \epsilon' \gamma_{\mu}]+\\&&
~~~~~~[\bar{\Psi}_P^{++}(\hat{q})(M-2\omega)I'_2+\bar{\Psi}_P^{--}(\hat{q})(M+2\omega)I''_2]\not \epsilon' \not P\gamma_{\mu}\bigg],
\end{eqnarray}

where $I_1',I_1'',I_2'$ and $I_2''$  are again the results of $Md\sigma$ integrations in the complex $\sigma$-plane over the poles of propagators, $S_F(\pm p_{1,2})$, and $S_F(p_1')$ in Fig.1 and are given in Eq.(57) in Appendix, with their 3D plots in Fig.2.

Here, $\bar{\Psi}^{\pm \pm}$ can be expressed as,

\begin{eqnarray}
&&\nonumber \bar{\Psi}_P^{++}=N_P\phi_P(\hat{q})[a_1'+ib_1'{\not}P+d_1'{\not}P{\not}q]\gamma_5\\&&
\bar{\Psi}_P^{--}=N_P\phi_P(\hat{q})[a_2'+ib_2'{\not}P+d_2'{\not}P{\not}q]\gamma_5,
\end{eqnarray}
and the coefficients associated with various Dirac structures are:

\begin{eqnarray}
&&\nonumber a_1'=M\frac{\omega+m}{2\omega}+\frac{m\hat{q}^2}{2\omega m};~ b_1'=-m\frac{\omega+m}{2\omega^2};\\&&
\nonumber d_1'=-\frac{\omega+m}{2\omega^2}+\frac{1}{4m}-\frac{m}{4\omega^2}+\frac{\hat{q}^2}{4\omega^2m}\\&&
\nonumber a_2'=M\frac{\omega-m}{2\omega}-\frac{m\hat{q}^2}{2\omega m};~ b_2'=m\frac{\omega-m}{2\omega^2};\\&&
d_2'=\frac{\omega-m}{2\omega^2}+\frac{1}{4m}-\frac{m}{4\omega^2}+\frac{\hat{q}^2}{4\omega^2m}.
\end{eqnarray}

with $m$ being the quark mass. Following a sequence of steps, we obtain the total amplitude for the process, $M_{fi}$, where both the diagrams in Fig.1 contribute equally as,
\begin{eqnarray}
&&\nonumber M_{fi}=[\bar{v}^{s_2}(p_2)\gamma_{\mu}u^{s_1}(p_1)]\beta \epsilon_{\mu\nu\alpha\beta}P_{\nu}\epsilon^{\lambda'}_{\alpha}k_{\beta};\\&&
\nonumber \beta=\frac{8e e_Q^2N_P}{M^4 s}\int\frac{d^3\hat{q}}{(2\pi)^3}X_2'\phi_P(\hat{q});\\&&
X_2'=b_1'(M-2\omega)I_1'+b_2'(M+2\omega)I_1''.
\end{eqnarray}

We can again express $M_{fi}=[\bar{v}^{s_2}(p_2)\gamma_{\mu}u^{s_1}(p_1)]<\gamma,\eta_c|J_{\mu}|0>$, where $<\gamma,\eta_c|J_{\mu}|0> = \beta \epsilon_{\mu\nu\alpha\beta}P_{\nu}\epsilon^{\lambda'}_{\alpha}k_{\beta}$, and is again gauge invariant. In amplitude, $M_{fi}$ for the process, $e^-e^+\rightarrow \gamma \eta_c$, there is a single form factor, $\beta(\sqrt{s})$ that absorbs the entire momentum integration over $d^3\hat{q}$. We again study the continuity structure of its integrand, $I_{\beta}(\hat{q},\sqrt{s})$, which is again a 2D function of both $\hat{q}$ and $\sqrt{s}$. We do a 3D plot of this integrand for $0<\hat{q}<3GeV$, and $4<\sqrt{s} <12$ GeV, which is given in Fig.6.

\begin{figure}[h!]
 \centering
 \includegraphics[width=15cm,height=8cm]{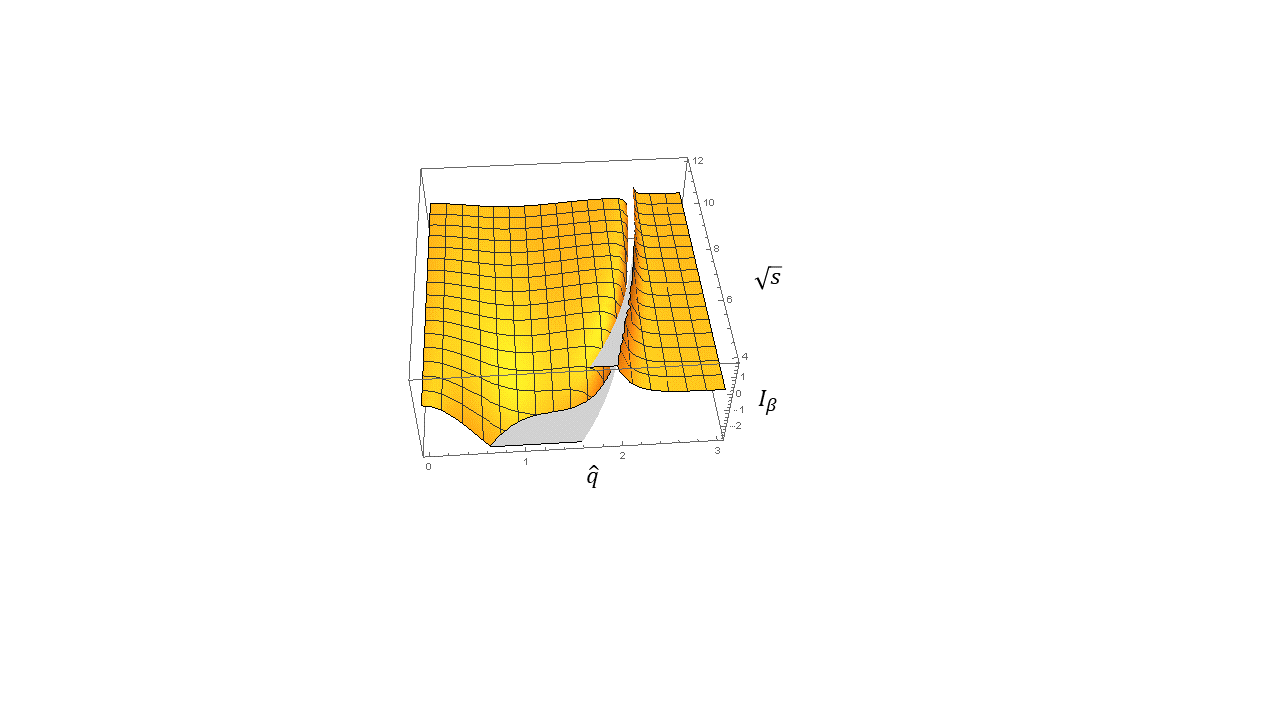}
 \caption{3D plots of integrand $I_{\beta}$ of form factor, $\beta$ as function of $\hat{q}$ and $\sqrt{s}$ for range of variables, $0< \hat{q} <3$ GeV, and $4 < \sqrt{s} < 12$ GeV for the process, $e^- e^+ \rightarrow \gamma \eta_{c}$.}
\end{figure}

 The plot shows a region of discontinuity, which has to be removed in evaluation of the $d^3\hat{q}$ integral in Eq.(53) for the form factor, $\beta$. To precisely identify the discontinuous region, we do a 2D plot of the integrand of $\beta$ versus $\hat{q}$ for each $\sqrt{s}$, as in Fig.4. The regions of discontinuity are identified up to two places of decimal for each $\sqrt{s}$. We list below these regions for some of the vales of $\sqrt{s}$ in Table 7.

\begin{table}[htp]
  \begin{center}
  \begin{tabular}{p{2cm}p{4cm}}
  \hline
 $\sqrt{s}$ &  $\hat{q}$ (for $\beta$) \\
  \hline
  12  &2.37$<\hat{q}<$ 2.41\\
  10.6&2.32$<\hat{q}<$2.36   \\
  8    &2.18$<\hat{q}<$ 2.22\\
  6    &1.97 $<\hat{q}<$ 2.01 \\
  4.6    &1.73$<\hat{q}<$1.77 \\
       \hline
  \end{tabular}
\caption{Regions of discontinuities in hadron internal momentum, $\hat{q}$(in GeV) in the integrand $I_{\beta}$ of form factor $\beta$, in Eq.(53) at different center of mass energies, $\sqrt{s}$ (in GeV) in the range, $4.6 - 12$ GeV for the process $e^- e^+ \rightarrow \gamma \eta_c(1P)$.}
\end{center}
\end{table}

The numerical value of $\beta(\sqrt{s})$ is given in Table 6 at $\sqrt{s}$ = 10.6 GeV and 4.6 GeV. Now, the spin averaged amplitude modulus squared, $|\bar{M}_{fi}|^2=\frac{1}{4}\sum_{s_1,s_2,\lambda'}|M_{fi}|^2$ can be expressed as,
\begin{equation}
|\bar{M}_{fi}|^2=\frac{1}{4}\beta^2[-2m^2P.kP.k+2p_1.PP.kp_2.k+2p_2.PP.kp_1.k+2p_1.kM^2p_2.k],
\end{equation}

which can be written as,
\begin{equation}
|\bar{M}_{fi}|^2=\frac{1}{4}\beta^2[16m_e^2M^4+4m_e^2s^2+s^3(1+cos^2\theta)-M^2(16m_e^2s+s^2(1+3cos^2\theta))]
\end{equation}
The cross section can be calculated using Eqs.(32), with $|\overrightarrow{P}'|=\frac{s-M^2}{\sqrt{s}}$ being the momentum of either of the final particles in the center of mass frame.  The BS normalizer for pseudoscalar meson behaves as, $N_P \sim M^{-2}$, while the form factor, $\beta \sim M^{-3}$, due to which the cross section again behaves as, $\sigma \sim M^{-2}$. The results of cross section for $e^- e^+\rightarrow \gamma \eta_c$ along with data and results of other models are given in Table 8, while the results of $N_P$ and $\beta$ are given in Table 9.

\begin{table}[hhhhhh]
  \begin{center}
  \begin{tabular}{p{3.5cm}p{1cm}p{1.5cm} p{4.5cm} p{1.1cm} p{1.5cm} p{2.0cm} }
  \hline
 Process                           &$\sqrt{s}$     & BSE& Experiment\cite{jia18,ablikim17}  &\cite{li13}& \cite{chung08}&\cite{braguta10}\\
 $e^- e^+\rightarrow \gamma \eta_c(1S)$&10.6&41.4575&$<21.1$ &70 &$82.0^{+21.4}_{-19.8}$& 41.6$\pm$14.1\\
 $e^- e^+\rightarrow \gamma \eta_c(1S)$&4.6&310.449&$(0.23\pm 0.53\pm 0.35)\times 10^{3}$ &678&& \\
 $e^- e^+\rightarrow \gamma \eta_c(1S)$&4.0&178.271&$(0.44\pm 1.02\pm 0.32)\times 10^{3}$ &540&& \\

    $e^- e^+\rightarrow \gamma\eta_c(2S)$&10.6&22.371&   &32 &$49.2^{+9.4}_{-7.4}$ &  24.2$\pm$14.5  \\
  $e^- e^+\rightarrow \gamma\eta_c(2S)$&4.6&19.702&   &33 &&   \\
       \hline
  \end{tabular}
\caption{Cross sections for processes, $e^- e^+\rightarrow \gamma \eta_{c} (nP)~ (n=1,2)$ (in fb) calculated in BSE at $\sqrt{s}$ = 10.6 4.6, and 4 GeV. along with recent data from Belle\cite{jia18}(at 10.58 GeV), BESIII\cite{ablikim21} (at 4.6 GeV, and 4 GeV), and other models}
\end{center}
\end{table}

\begin{table}[hhh]
  \begin{center}
  \begin{tabular}{p{3.5cm}p{2cm} p{1.9cm} p{2.9cm} }
  \hline
 Process& $\sqrt{s}$&                                $N_P$ & $\beta$\\
  \hline
  $e^- e^+\rightarrow \gamma \eta_c (1S)$&10.6&4.545   &$-8.3977\times 10^{-7}$  \\
  $e^- e^+\rightarrow \gamma \eta_c(2S)$&10.6&4.388   & $6.873\times 10^{-7}$ \\
  $e^- e^+\rightarrow \gamma \eta_c (1S)$&4.6&4.545   &$-2.4989\times 10^{-5}$  \\
  $e^- e^+\rightarrow \gamma \eta_c(2S)$&4.6&4.388   & $1.915\times 10^{-5}$ \\
       \hline
  \end{tabular}
\caption{Numerical values of BS normalizer, $N_P$ (in $GeV^{-2}$) for $\eta_c(1S)$, and $\eta_c(2S)$, along with the numerical values of $\beta$ (in $GeV.^{-3}$) for the processes, $e^- e^+\rightarrow \gamma \eta_c(1S)$, and $e^- e^+\rightarrow \gamma \eta_c(2S)$ at $\sqrt{s}=10.6 GeV$, and $4.6 GeV$}
\end{center}
\end{table}

\begin{figure}[h!]
 \centering
 \includegraphics[width=20cm,height=10cm]{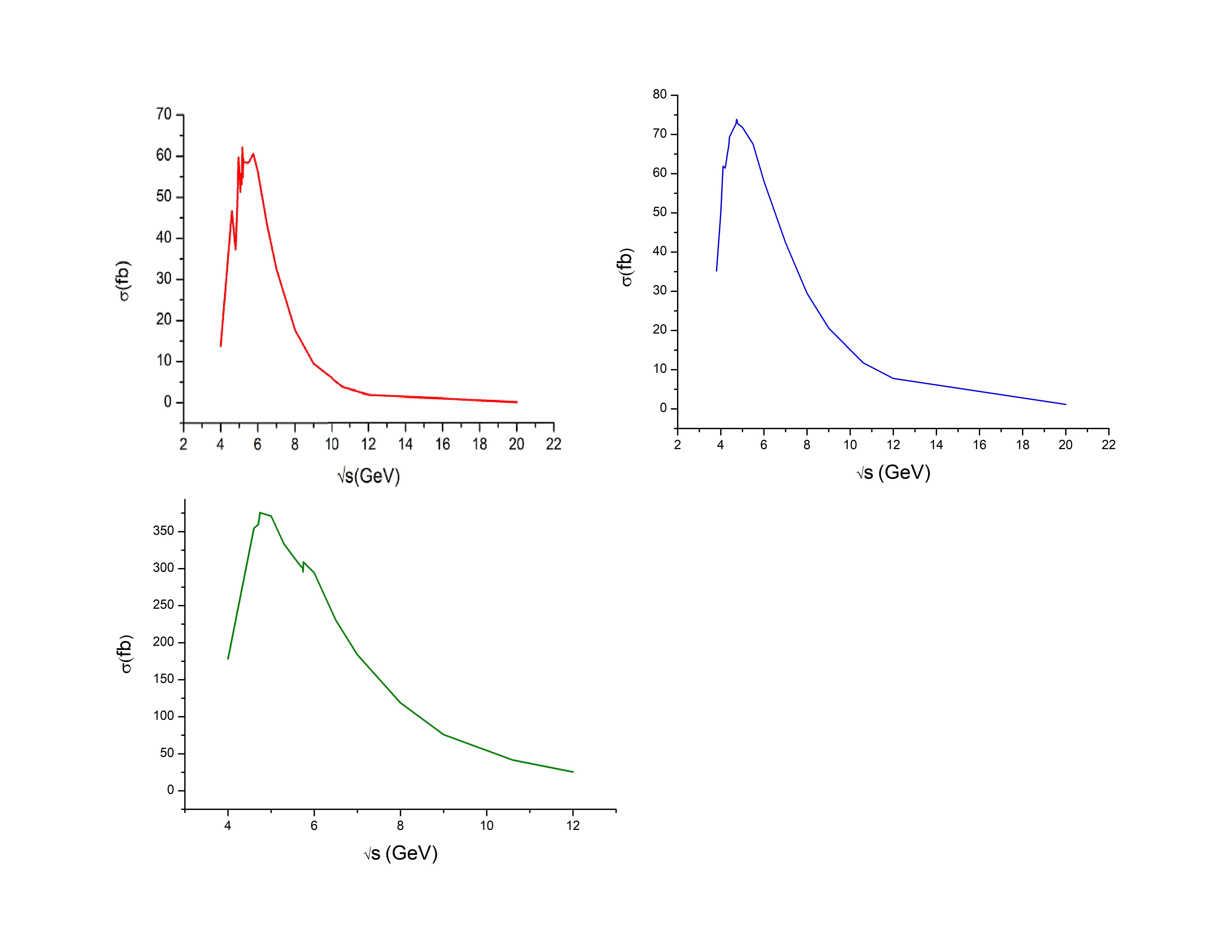}
 \caption{Plots of the cross section $\sigma$ (in fb) versus $\sqrt{s}$ (in GeV) for the processes, $e^- e^+ \rightarrow \gamma\rightarrow \gamma \chi_{c0}$}, $e^- e^+ \rightarrow \gamma \chi_{c1}$ (in first line), and $e^- e^+ \rightarrow \gamma \eta_c$ (in second line) for energy range, $\sqrt{s}$ = 4 - 20 GeV.
\end{figure}

\section{Discussions}
In this work we study the  production of ground and excited charmonium states in $e^- e^+ \rightarrow \gamma \rightarrow \gamma \chi_{cJ}(nP)$ for $J=0,1$, and $e^- e^+ \rightarrow \gamma \rightarrow \gamma \eta_c$ through leading order (LO) diagrams, which proceed through exchange of a virtual photon that is coupled to the $\gamma$ and $\chi_{cJ}$ through  triangle quark loop diagram, in the framework of $4\times 4$ Bethe-Salpeter equation under Covariant Instantaneous Ansatz, at center of mass energies, $\sqrt{s}=10.6$ GeV (Belle energy), and $4.6$ GeV (BESIII energy), and calculate the cross sections for these processes. Plots of cross sections versus the center-of mass energy, $\sqrt{s}$ reveal small fluctuations in their behaviour for all the three processes in the low energy ($\sqrt{s}$ = 4 - 6 GeV) region as seen from Fig.7 and are analyzed in terms of the form factors. Such fluctuations in the behaviour of cross sections for all the three processes at BESIII energies are also observed in \cite{ablikim21,ablikim17}. The numerical results of cross sections obtained for the three processes at $\sqrt{s}$=10.6 GeV, 4.6 GeV and 4 GeV, are compared with the only available experimental results for $e^-e^+\rightarrow \gamma \rightarrow \gamma H$ ($H=\chi_{c0,c1},\eta_c$) at 10.6 GeV (measured by Belle collaboration\cite{jia18}), and at 4.6, and 4 GeV (measured by BESIII collaboration\cite{ablikim21}), as well as other models \cite{braguta10,brambilla20, chung21,sang20}.

Now, as regards the details of cross sections in our model is concerned, for the process, $e^- e^+\rightarrow \gamma \chi_{c0}$, at $\sqrt{s}=10.6$ GeV, we obtain the cross section, $\sigma=3.81$ fb, which is consistent with the upper limit of $\sigma$ estimates by the Belle collaboration at $< 205.9$ fb \cite{jia18}, and can also be compared with the corresponding cross sections of this process obtained in Refs.[18, 19, 47, 48, 51]. Similarly at 4.6 GeV, our cross section for this process is 46.617 fb, which is  consistent with the upper limit of $\sigma <2.6\times 10^{3}$ fb measured recently by BESIII collaboration\cite{ablikim21}. And at 4 GeV, we obtain cross section that is smaller than the cross section at 4.6 GeV, but it is again consistent with the upper limit of $\sigma < 4.5\times 10^3$ fb. Thus in our calculations of this process, we find that, $\sigma$ (at 10.6 GeV)$ < \sigma$(at 4 GeV) $<\sigma$ (at 4.6 GeV), though in Ref.[51], $\sigma$(at 10.6 GeV) $< \sigma$ (at 4.6 GeV) $< \sigma$ (at 4 GeV) for this process. However, our cross sections (at 4.6 GeV, and 4 GeV) are consistent with the experimental upper limits for this process listed in \cite{ablikim21}

Similarly, for $e^- e^+\rightarrow \gamma \rightarrow \gamma \chi_{c1}(1P)$, we obtain the cross section, $\sigma=12.071fb$, at 10.6 GeV, which is consistent with Belle measurement of $17.3^{+4.2}_{-3.9}\pm1.7$ fb\cite{jia18} within $1.2\sigma$, and can be compared with results of [18, 19, 47, 48, 51], while at 4.6 GeV, our cross section for this process is 71.551 fb, which is within $2\sigma$ of $(1.7^{+0.8}_{-0.6}\pm0.2)\times 10^{3}$ measured recently by BESIII\cite{ablikim21}. And at 4 GeV, we obtain $\sigma$ = 50.055 fb, which is around 2.8$\sigma$ of BESIII cross section $(4.5^{+1.5}_{-1.3}\pm0.4)\times 10^{3}$ fb\cite{ablikim21}. Thus, for this process, we again obtain $\sigma$ (10.6 GeV) $< \sigma$ (4 GeV) $< \sigma$ (4.6 GeV), which is consistent with the trend in cross section values in [51].

Similarly for $e^- e^+\rightarrow \gamma \eta_c(1P)$, at 10.6 GeV, as checked from Table 5, we obtain $\sigma$ = 41.4575 fb, which is higher than the upper limit of Belle measurement of $<21.1$ fb\cite{jia18}, but can be compared with results of [48, 50, 51]. Further at BESIII energies, we obtain $\sigma$ (at 4.6 GeV) = 310.449 fb (c.f. $\sigma =(0.23\pm 0.53\pm 0.35)\times 10^{3}$ fb \cite{ablikim17}), and $\sigma$ (at 4 GeV) =178.271 fb (c.f. $(0.44\pm 1.02\pm 0.32)\times 10^{3}$ fb\cite{ablikim17}), where both these measurements are within 1$\sigma$ of experimental data\cite{ablikim17}. And again, we find the trend, $\sigma$ (10.6 GeV) $< \sigma$ (4 GeV)$ < \sigma$ (4.6 GeV), which is also consistent the trend seen in cross sections for this process in Ref.[51].

Thus, our theoretical results of cross sections at 10.6 GeV are in good agreement with experiment\cite{jia18}, and other models for all the three processes. But in the low energy BESIII region (4.6 - 4 GeV), though our cross sections for $(\gamma \chi_{c0})$, and $(\gamma \eta_c)$ are in good agreement with data\cite{ablikim21,ablikim17}, however our cross section for $(\gamma \chi_{c1})$ production deviates from available experimental results by $2\sigma$ (for 4.6 GeV), and  2.8$\sigma$ (for 4 GeV). With more detailed calculations in BSE involving higher order QCD corrections in this framework, which is beyond the scope of this paper, we expect this discrepancy can be resolved. Such higher order QCD corrections in NRQCD have been attempted recently in Ref.\cite{feng19}.

We have also drawn plots of $\sigma$ (in fb) for each of these three processes in Fig.7, as a function of the center of mass energy, $\sqrt{s}$ in the range $4 - 20$ GeV. We also obtain minor fluctuations in cross sections for all the three processes, that can be analyzed in terms of their form factors. Such fluctuations in cross sections in BESIII region (4.6 - 4 GeV) have also been observed in \cite{ablikim21,ablikim17}.

For the process $e^-e^+\rightarrow \gamma \rightarrow \gamma \chi_{c0}$, with decrease in energy, the plot of cross section shows first a continuous rise followed by fall to a value $46.667$ fb at 4.6 GeV, which is consistent with the upper limit estimates of BESIII\cite{ablikim21} collaboration. However, from our plot of $\sigma$, we also notice small fluctuations of order 5 -10 fb, in the low energy region 6 - 4 GeV.  To analyze this, in Fig. 8, we have further drawn the plots of the form factors $\beta_1$ and $\beta_3$ (versus $\sqrt{s}$) in the energy region 12 GeV - 4 GeV, along with the plots of parts of cross sections, $\sigma_A$, and $\sigma_B$ (where $\sigma=\sigma_A +\sigma_B$), each of which can be associated with the squares of the form factors $\beta_{1}^2$ and $\beta_{2}^2$ respectively, arising due to splitting of expression Eq.(30) in $|\bar{M}_{fi}|^2|$ into two parts.

\begin{figure}[h!]
 \centering
 \includegraphics[width=20cm,height=10cm]{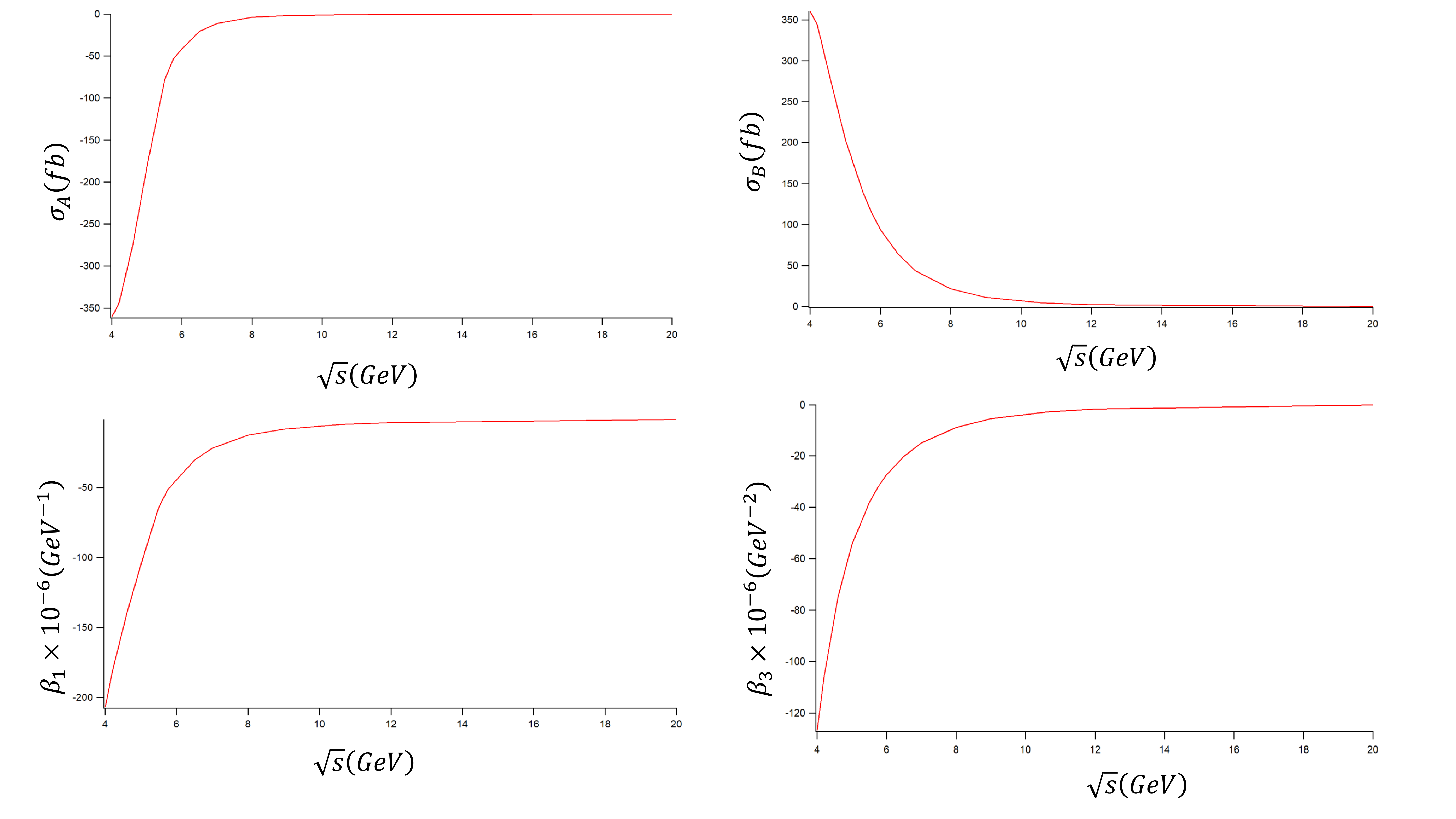}
 \caption{Line 1 represents the plots of the cross section $\sigma_A$ (associated with form factor, $\beta_1$), $\sigma_B$(associated with form factor, $\beta_3$) in Eqs.(29), versus $\sqrt{s}$ (in GeV) for the process, $e^- e^+ \rightarrow \gamma\rightarrow \gamma \chi_{c0}$ for energy range, $\sqrt{s}$ = 4 - 20 GeV. Line 2 represents the plots of form factors $\beta_1$(in $GeV^{-1}$) and $\beta_3$ (in $GeV^{-2}$), expressible as integrals in Eq.(25) versus $\sqrt{s}$ (in GeV).}
\end{figure}

The plot of $\beta_3$ is quite smooth, but the plot of $\beta_1$, shows minor variations in low energy region. However, since the cross sections $\sigma_A$ and $\sigma_B$ involve squares of the form factors, $\beta_1^2$ and $\beta_3^2$, the minor variations that exist in $\beta_1$ get amplified leading to fluctuations in $\sigma_A$. And any fluctuation in total cross section, $\sigma$ in any given energy sub-interval in the low energy region 6 - 4 GeV (as shown in the cross section plot of this process in Fig.7), can be explained due to different rates of variations of $\sigma_A$, and $\sigma_B$ with change in $\sqrt{s}$ in that interval.

For the process, $e^- e^+\rightarrow \gamma \chi_{c1}$, we again see some minor fluctuations in the cross section plot in Fig.7, which can be explained by splitting the cross section, $\sigma=\sigma_C + \sigma_D$ (through splitting $|\bar{M}_{fi}|^{2}$ in Eq.(45)) for the above process, where $\sigma_C$ involves, $|\bar{M}_{fi}|_{C}^{2}$ (associated with the square of the form factor, $g_{1}^2$), while, $\sigma_D$ involves, $|\bar{M}_{fi}|_{D}^{2}$ (associated with $g_{2}^2$). We also plot $\sigma_C$ and $\sigma_D$ versus $\sqrt{s}$ in Fig.9 for energies $\sqrt{s}$ in region, 4- 20 GeV.

The behaviour of total cross section, $\sigma$ for the above process is very similar to the behaviour of $\sigma_C$ for this process, due to the percentage contribution of $\sigma_D$ to total cross section being negligible, ranging between $1.304\% - 1.670\%$ in the region, $4.0 < \sqrt{s} < 6.0$ GeV. And though $\sigma_D$ in Fig.9 fluctuates rapidly in the low  energy region, 5.5 - 4 GeV (due to variation in form factor, $g_3$ in this region), the magnitude of variation of $\sigma_D$ in comparison to the variation of $\sigma_C$ in any given energy sub-interval is negligible. Thus the behaviour of total cross section for the above process is mainly due to the behaviour of $\sigma_C$, which can also be seen from the similarity of the graphs of $\sigma_C$, and $\sigma$.

\begin{figure}[h!]
 \centering
 \includegraphics[width=20cm,height=10cm]{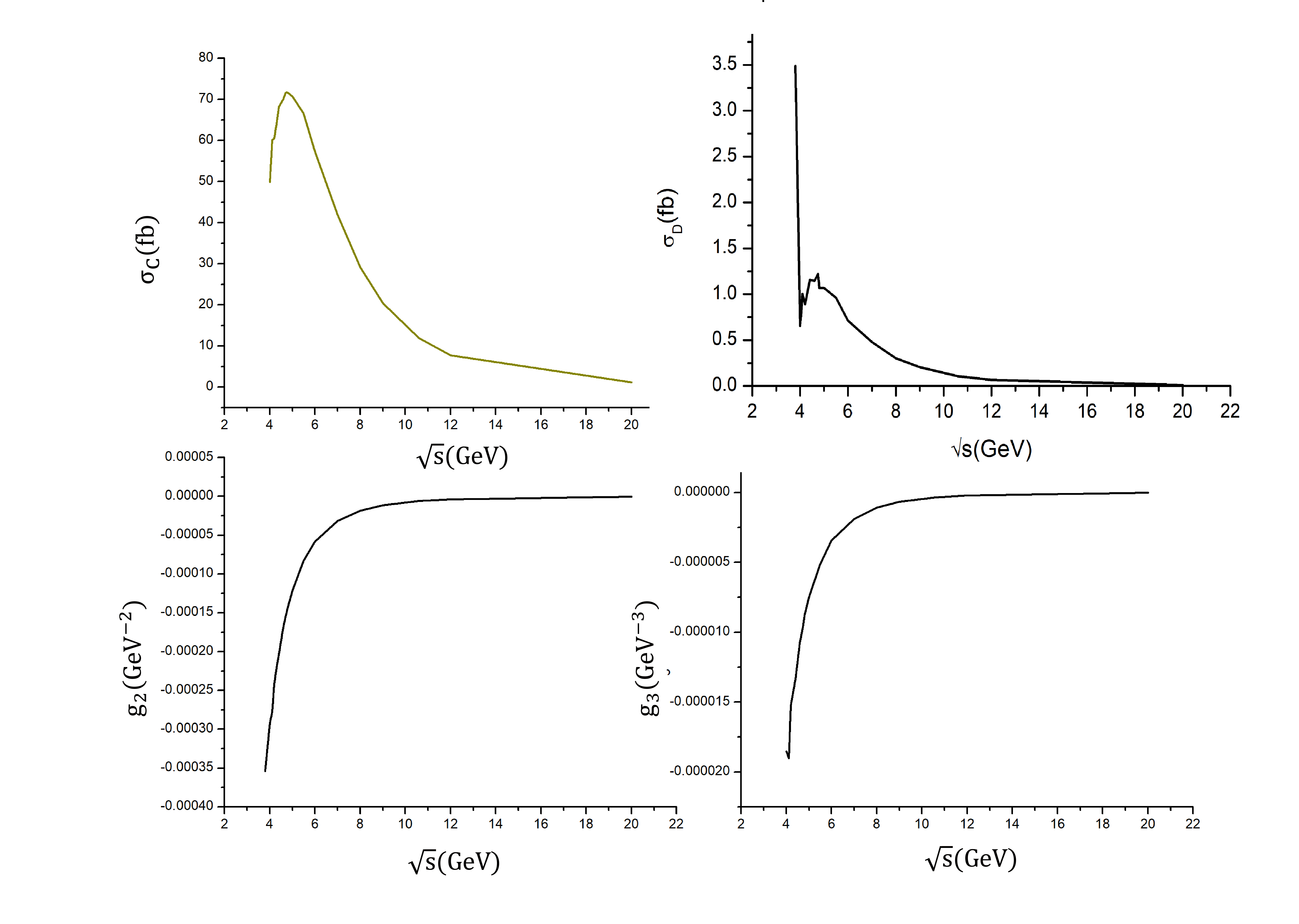}
 \caption{Line 1 represents the plots of the cross section $\sigma_C$ (associated with form factor, $g_2$) and $\sigma_D$(associated with form factor, $g_3$) in Eqs.(32) and (45)), versus $\sqrt{s}$ (in GeV.) for the process, $e^- e^+ \rightarrow \gamma\rightarrow \gamma \chi_{c1}$ for energy range, $\sqrt{s}$ = 4 - 20 GeV. Line 2 represents the plots of form factors $g_2$(in $GeV^{-2}$) and $g_3$ (in $GeV^{-3}$), expressible as integrals in Eq.(40) versus $\sqrt{s}$ (in GeV).}
\end{figure}

\begin{figure}[h!]
 \centering
 \includegraphics[width=20cm,height=10cm]{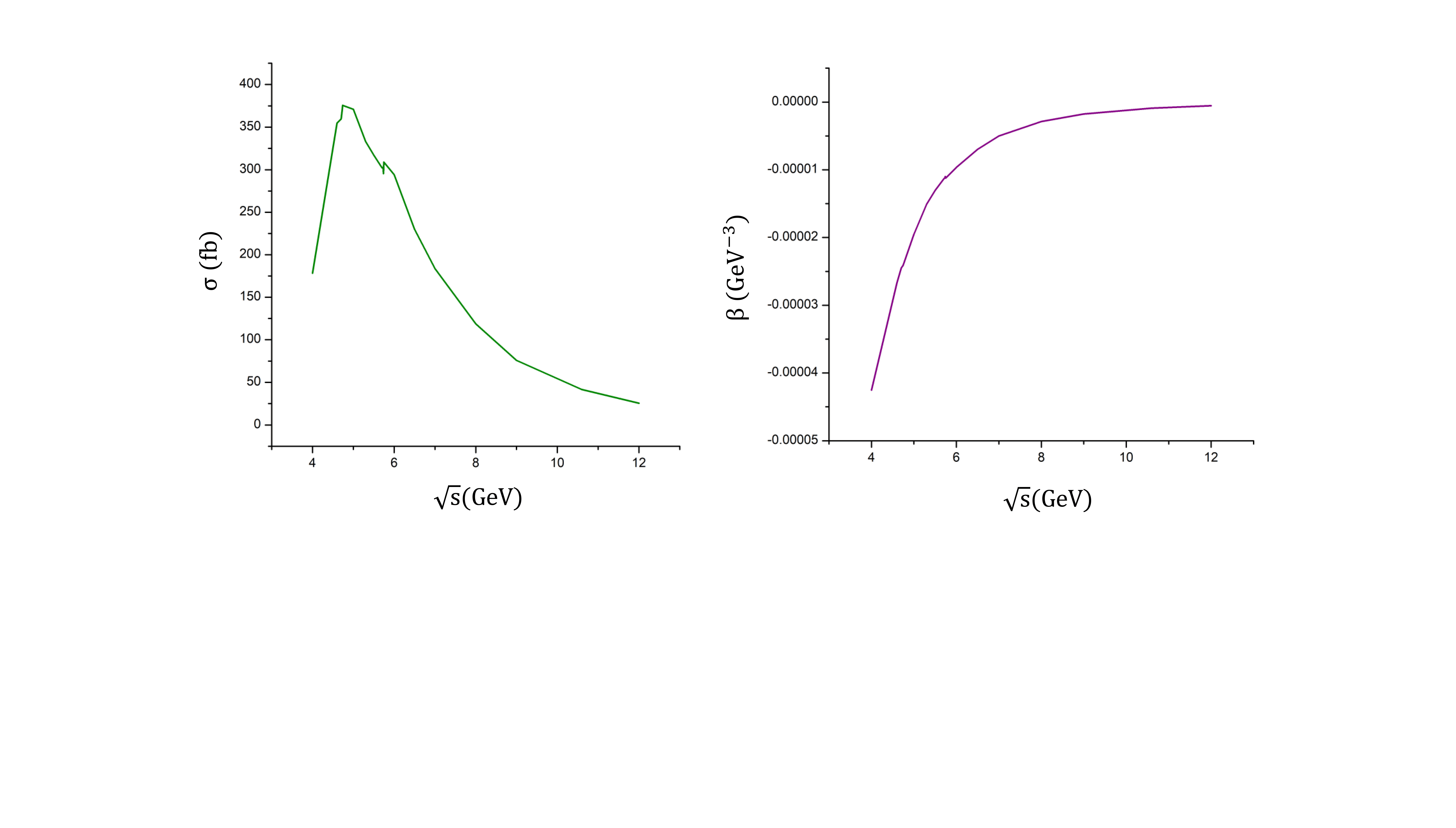}
 \caption{Line 1 represents the plot of the cross section $\sigma$ , along with the plot of the form factor $\beta$ versus $\sqrt{s}$ (in GeV.) for the process, $e^- e^+ \rightarrow \gamma\rightarrow \gamma \eta_{c}$ for energy range, $\sqrt{s}$ = 4 - 20 GeV.}
\end{figure}

Regarding the process, $e^- e^+\rightarrow \gamma \eta_c$, there is again a continuous rise in cross section followed by a fall with decrease in center of mass energy from $\sqrt{s}$=10.6GeV to 4 GeV (where $\sigma$=178.271fb). Also, there are small fluctuations in cross section between 5.74GeV and 4.74GeV, to analyse which we draw the plot of the form factor $\beta$, which is a smooth plot with a continuous fall from 12 GeV to 4GeV, except for small kinks at $\sqrt{s}=$ 5.74GeV and 4.74GeV, and these are precisely the points where one sees small fluctuations in cross section. This might be due to the reason that the minor kinks in plot of $\beta$ at 5.74GeV, and 4.74GeV get amplified due to the presence of $\beta^2$ in $|\bar{M}_{fi}|^2$, leading to minor fluctuations in cross section $\sigma$ at these very points. However, the fluctuations in cross section values shown in our cross sections in lower energy region (4- 6 GeV), is consistent with the fluctuations in BESIII\cite{ablikim21,ablikim17} data in the lower energy region $\sqrt{s}$= 4 - 6 GeV, which might show some interesting physics.

We have also calculated the cross sections for $e^-e^+ \rightarrow \gamma \chi_{c0,c1}(2P)$, and $e^-e^+\rightarrow \gamma \eta_c(2S)$ at 10.6 GeV for which data is not yet available. However our results of cross sections with leading order diagrams alone for all these processes provide a sizable contribution, which might be mainly due to the BSE being a fully relativistic approach that incorporates the relativistic effect of quark spins and can also describe internal motion of constituent quarks within the hadron in a relativistically covariant manner. We wish to point out that the results of our study using BSE is further validated by a recent calculation \cite{luchinsky} on $e^- e^+ \rightarrow J/\Psi \eta_c$ using Light cone expansion, where the authors had shown that by taking intrinsic motion of quarks inside the hadrons, one can significantly increase the value of cross section.

Further calculations are being done on incorporation of higher order QCD corrections to these processes in the framework of BSE, which is expected to improve the results further, but is beyond the scope of this paper. This calculation involving production of $\gamma \chi_{cJ}$, and $ \gamma \eta_c$ in electron-positron annihilation can be easily extended to studies on other processes (involving the exchange of a single virtual photon) observed at B-factories such as,$e^+ e^- \rightarrow H' H''$ on lines of the work done in the present paper. We further wish to extend this work to studies on processes involving $J/\psi+J/\psi$ production in electron-positron annihilation proceeding through two intermediate photons.

\section{Appendix}
The results of $Md\sigma$ integrations, $I_1',I_1'',I_2'$ and $I_2''$ (in Eqs.(20),(38) and (50)) in the complex $\sigma$-plane over the poles of propagators, $S_F(\pm p_{1,2})$, and $S_F(p_1')$ in Fig.1, for $e^-e^+\rightarrow \gamma \chi_{cJ}(J=0,1)$, and $e^-e^+\rightarrow \gamma \eta_c$ are given as,

\begin{eqnarray}
&&\nonumber I'_1=\int \frac{Md\sigma}{2\pi i}\frac{1}{[\sigma-(-\frac{1}{2}+\frac{\omega}{M})][\sigma-(\frac{1}{2}-\frac{\omega}{M})][\sigma-(-\frac{1}{2}-\frac{2E^2}{M^2}+\frac{1}{M}\sqrt{\omega^2+\frac{4E^4}{M^2}})][\sigma-(-\frac{1}{2}-\frac{2E^2}{M^2}-\frac{1}{M}\sqrt{\omega^2+\frac{4E^4}{M^2}})]}\\&&
\nonumber ~~~~~~~~~~~~=\frac{M^6[M-2(\omega+\Lambda)]}{2(M-2\omega)\Lambda[2E^2+M(\omega+\Lambda)][-2E^2+M(-M+\omega+\Lambda)]};\\&&
\nonumber I''_1=\int \frac{Md\sigma}{2\pi i}\frac{1}{[\sigma-(-\frac{1}{2}-\frac{\omega}{M})][\sigma-(\frac{1}{2}+\frac{\omega}{M})][\sigma-(-\frac{1}{2}-\frac{2E^2}{M^2}+\frac{1}{M}\sqrt{\omega^2+\frac{4E^4}{M^2}})][\sigma-(-\frac{1}{2}-\frac{2E^2}{M^2}-\frac{1}{M}\sqrt{\omega^2+\frac{4E^4}{M^2}})]}\\&&
\nonumber~~~~~~~~~~~~~~=\frac{M^6[M+2(\omega+\Lambda)]}{2(M+2\omega)\Lambda[-2E^2+M(\omega+\Lambda)][2E^2+M(M+\omega+\Lambda)]};\\&&
\nonumber I'_2=\int \frac{Md\sigma}{2\pi i}\frac{\sigma}{[\sigma-(-\frac{1}{2}+\frac{\omega}{M})][\sigma-(\frac{1}{2}-\frac{\omega}{M})][\sigma-(-\frac{1}{2}-\frac{2E^2}{M^2}+\frac{1}{M}\sqrt{\omega^2+\frac{4E^4}{M^2}})][\sigma-(-\frac{1}{2}-\frac{2E^2}{M^2}-\frac{1}{M}\sqrt{\omega^2+\frac{4E^4}{M^2}})]}\\&&
\nonumber ~~~~~~~~~~~=\frac{M^4(4E^2+M^2)}{4\Lambda[2E^2+M(\omega+\Lambda)][-2E^2+M(-M+\omega+\Lambda)]};\\&&
\nonumber I''_2=\int \frac{Md\sigma}{2\pi i}\frac{\sigma}{[\sigma-(-\frac{1}{2}-\frac{\omega}{M})][\sigma-(\frac{1}{2}+\frac{\omega}{M})][\sigma-(-\frac{1}{2}-\frac{2E^2}{M^2}+\frac{1}{M}\sqrt{\omega^2+\frac{4E^4}{M^2}})][\sigma-(-\frac{1}{2}-\frac{2E^2}{M^2}-\frac{1}{M}\sqrt{\omega^2+\frac{4E^4}{M^2}})]}\\&&
\nonumber~~~~~~~~~~=\frac{M^4(4E^2+M^2)}{4\Lambda[-2E^2+M(\omega+\Lambda)][2E^2+M(M+\omega+\Lambda)]},\\&&
\Lambda=\sqrt{\frac{4E^4}{M^2}+\omega^2}
\end{eqnarray}

\end{document}